\documentclass[%
 aip,
 amsmath,amssymb,
 reprint,%
]{revtex4-1}

\usepackage{graphicx}% Include figure files
\usepackage{dcolumn}% Align table columns on decimal point
\usepackage{bm}% bold math
\usepackage[utf8]{inputenc}
\usepackage[T1]{fontenc}
\usepackage{mathptmx}
\usepackage{etoolbox}
\usepackage{float} % added by Jonah. For adding [H] modifier to figures
\usepackage{verbatim} % added by Jonah. For adding block comments
\usepackage{wrapfig} % added by Jonah

\makeatletter
\def\@email#1#2{%
 \endgroup
 \patchcmd{\titleblock@produce}
  {\frontmatter@RRAPformat}
  {\frontmatter@RRAPformat{\produce@RRAP{*#1\href{mailto:#2}{#2}}}\frontmatter@RRAPformat}
  {}{}
}%
\makeatother

\begin{document}

\preprint{AIP/123-QED}

\title{Coriolis Force Compensation and Laser Beam Delivery for 100-Meter Baseline Atom Interferometry}

%{Coriolis Force Compensation and Laser Beam Pointing Control for MAGIS-100}

%could say "for MAGIS-100" -- although this could be a more general method
% we could make the title more general, and then mention MAGIS-100 in the intro
% or we could put MAGIS in the title and say this could apply to other systems

\author{Jonah Glick}
\affiliation{Department of Physics and Astronomy and Center for Fundamental Physics, Northwestern University, Evanston, Illinois 60208, USA}

\author{Zilin Chen}
\affiliation{Department of Physics and Astronomy and Center for Fundamental Physics, Northwestern University, Evanston, Illinois 60208, USA}

\author{Tejas Deshpande}
\affiliation{Department of Physics and Astronomy and Center for Fundamental Physics, Northwestern University, Evanston, Illinois 60208, USA}

\author{Yiping Wang}
\affiliation{Department of Physics and Astronomy and Center for Fundamental Physics, Northwestern University, Evanston, Illinois 60208, USA}

\author{Tim Kovachy}
\affiliation{Department of Physics and Astronomy and Center for Fundamental Physics, Northwestern University, Evanston, Illinois 60208, USA}
\email{timothy.kovachy@northwestern.edu}

\date{\today}% It is always \today, today,

\begin{abstract}
The Coriolis force is a significant source of systematic phase errors and dephasing in atom interferometry and is often compensated by counter-rotating the interferometry laser beam against Earth's rotation. We present a novel method for performing Coriolis force compensation for long-baseline atom interferometry which mitigates atom-beam misalignment due to beam rotation, an effect which is magnified by the long lever arm of the baseline length.
%which mitigates the effect of beam rotation resulting in atom-beam misalignment.
%a technical challenge inherent to larger baselines--namely, that beam rotation can result in atom-beam misalignment that is magnified by the long lever arm provided by the baseline. 
The method involves adjustment of the angle of the interferometer beam prior to a magnifying telescope, enabling the beam to pivot around a tunable position along the interferometer baseline. By tuning the initial atom kinematics, and adjusting the angle with which the interferometer beam pivots about this point, we can ensure that the atoms align with the center of the beam during the atom optics laser pulses. This approach will be used in the MAGIS-100 atom interferometer and could also be applied to other long-baseline atom interferometers. An additional challenge associated with long baseline interferometry is that since long-baseline atom interferometers are often located outside of typical laboratory environments, facilities constraints may require lasers to be housed in a climate-controlled room a significant distance away from the main experiment. Nonlinear effects in optical fibers restrict the use of fiber-based transport of the high-power interferometry beam from the laser room to the experiment. We present the design of and prototype data from a laser transport system for MAGIS-100 that maintains robustness against alignment drifts despite the absence of a long fiber.

\end{abstract}

\maketitle

\section{\label{sec:introduction}Introduction}

% overview of atom interferometry -- with emphasis on long baselines
Atom interferometry is a powerful tool for testing fundamental physics \cite{fray_2004, PhysRevA.88.043615, Bouchendira_2011, PhysRevLett.113.023005, Schlippert_2014, kuhn_2014, rosi_2014, barrett_2015, Biedermann_2015, Zhou_2015, Barrett_2016, williams_2016, rosi_2017, Asenbaum_2017, Overstreet_2018, parker_2018, Asenbaum_2020, morel_2020, overstreet_2022, Hamilton_2015}. Atom intererometers with 100-meter-scale baselines and larger provide an opportunity to detect gravitational waves in a frequency range between 0.01 Hz and 3 Hz  \cite{Dimopoulos_2008, Graham_2013, Graham_2016, PhysRevD.93.021101, hogan_2011, abou_2020, MAGIS-100:2021etm}, search for ultra-light, wave-like dark matter\cite{Arvanitaki_2018, Graham_2016_2,Banerjee_2022}, and test quantum mechanics over macroscopic spatio-temporal scales\cite{kovachy_2015,MAGIS-100:2021etm}. Excitement about the potential scientific utility of long-baseline atom interferometry has spurred the ongoing construction of  multiple long-baseline atom interferometers, including AION\cite{Badurina_2020}, MIGA\cite{Canuel_2018}, ZAIGA \cite{Zhan_2019}, VLBAI\cite{Hartwig_2015}, and MAGIS-100\cite{MAGIS-100:2021etm}. The sensitivity of an atom interferometer generally increases with an increasing baseline length, but larger baselines introduce new technical challenges. In this paper, we address two key challenges associated with scaling up the baseline of an atom interferometer: 1. Traditional methods of compensating for systematic effects associated with the rotation of the Earth break down for longer baselines and 2. Delivering the interferometer beam from the climate-controlled laser room to the main experiment with alignment that is robust against drifts can be more difficult in areas such as deep shafts because of facilities constraints.
%Long-baseline atom interferometers may be located in areas such as deep shafts where facilities constraints complicate the ability to deliver the interferometery laser beam from a climate-controlled laser room to the interferometer with alignment that is robust against drifts. 
For instance, this is a major challenge faced by MAGIS-100.

%2. Longer baselines impose stricter requirements on the mechanical stability of the optical systems which deliver the interferometer beam to the atoms.

%problem with corilis forces, and existing solutions
Coriolis forces caused by the rotation of the Earth induce velocity dependent phase shifts in an atom interferometer \cite{dickerson_2013, PhysRevLett.108.090402}. Long baselines can enable long temporal spacing $T$ between beam splitter and mirror pulses, as well as large momentum transfer (LMT), but the scale of the Coriolis induced phase shifts increases with increasing $T$ and LMT number $n$. These phase shifts can wash out the interferometer contrast when the signal is extracted by averaging over the atomic ensemble. One solution to mitigate this effect is that in the limit in which the atom cloud can be approximated as expanding from a point source, the Coriolis phase shifts result in spatial fringes across the cloud which can be measured with spatially resolved detection \cite{dickerson_2013}. Even in this limit, however, there are two challenges posed by these velocity dependent phase shifts: 1. the Coriolis force can cause shot-to-shot fluctuations in the average initial kinematics of the cloud (e.g., from jitter in the atom cooling and launching processes) to manifest as fluctuating shot-to-shot overall phase shifts, and 2. As $T$ and $n$ increase, even in the point source limit, fringes could be of such a high spatial frequency that they are difficult to spatially resolve in detection. It is therefore important to adopt strategies to compensate for Coriolis forces.

%new coriolis compensation method for longer baselines
A method to compensate for Coriolis forces in atom interferometers has been successfully demonstrated and consists of rotating the interferometer beam against the rotation of the Earth with a single piezo-actuated mirror \cite{dickerson_2013, PhysRevLett.108.090402}. This scheme breaks down for longer interferometer baselines because the long lever arm between the piezo-actuated mirror and the atoms can cause the center of the interferometer beam to transversely shift away from the atom cloud as the beam is rotated.
%This scheme breaks down for longer interferometer baselines because beam rotations manifest as larger translations of the beam and can cause the center of the interferometer beam to be offset from the center of the atom cloud.
In section \ref{sec:pivot_point_method}, we present a new method to allow for Coriolis force compensation in longer baseline atom interferometers. This method keeps the atoms in the center of the interferometer beam while the beam is pulsed on and performing beam splitter or mirror atom optics operations. We also consider challenges associated with using this method to do Coriolis force compensation for km-scale baseline interferometers.

Additionally, while atom interferometers have to date often been located inside single rooms in a laboratory, 100-meter-scale and larger atom interferometers will tend to require facilities of a much larger scale such as deep shafts. Building these facilities can be cost prohibitive, but adapting existing facilities is more feasible.
%Such facilities are highly expensive to build, favoring the adaptation of existing facilities. 
Existing facilities of this type were typically not built with atom interferometry in mind and can therefore pose constraints that need to be overcome. In MAGIS-100, for example, the atom interferometry lasers will be located in a dedicated laser room to ensure stable environmental temperatures and to meet laser safety requirements, but space constraints require this laser room to be located approximately 10\;m from the shaft that contains the atom interferometer.  For distances of this scale, stimulated Brillouin scattering \cite{Kobyakov:10} significantly complicates the prospects for using a single-mode optical fiber to transport the high-power interferometry laser beam from the laser room to the shaft \cite{MAGIS-100:2021etm}.

A key advantage of fiber-based transport is that the position and pointing of the post-transport beam is determined solely by the fiber, which can be anchored to a structurally stable floor. On the other hand, free-space beam delivery is more susceptible to alignment drifts, especially with a relatively long $\approx 10\;$m lever arm.
%MOVED TO FURTHER DOWN IN PAPER -- (of course, stability of the alignment of the final delivery optics before the 100\;m baseline is also critical, but this is aided by the presence of a structurally stable floor and wall by the shaft--and by the fact that the final telescope that magnifies the size of the beam by a factor of 30 also demagnifies beam angles by the same factor, reducing the impact of angular drifts from pre-telescope optics).
In Section \ref{sec:beam_delivery_system}, we discuss the design of a laser transport system for MAGIS-100 which uses a combination of stable mechanical mounts, a relay imaging system, and a short fiber to provide delivery of the interferometer beam to the shaft that maintains the aforementioned benefits of delivery via a long fiber but avoids the limitations of stimulated Brillouin scattering. In Section \ref{sec:experimental_test_horizontal_config} we present data from a prototype test of this system. The stability of the alignment of the final delivery optics before the 100\;m baseline is also critical, but this is aided by the presence of a structurally stable floor and wall by the shaft, and by the fact that the telescope which magnifies the size of the beam also demagnifies beam angles, reducing the impact of angular drifts from pre-telescope optics.

\section{\label{sec:pivot_point_method} Pivot Point Method for Rotation Compensation}

\subsection{\label{sec:outline_of_pivot_point_method} Outline of the method}

%paragraph describing the problem
It has been demonstrated in a 10\;m baseline atom interferometer that counter-rotating an interferometer beam against the rotation of the Earth suppresses interferometer phase shifts and contrast loss from the coupling of Coriolis forces to the initial kinematics of the atom cloud \cite{dickerson_2013, PhysRevLett.108.090402,PhysRevLett.125.191101, PhysRevLett.111.113002}. Traditionally, this counter-rotation scheme has been done with a single piezo-actuated tip-tilt mirror at one end of the baseline, but this approach breaks down for larger interferomter baselines. The location of this mirror sets the position about which the interferometer beam pivots.  For larger baseline interferometers, the increased lever arm can cause the center of the interferometer beam to move away from the center of the atom cloud as it is rotated, which in the extreme case causes the beam to miss the atom cloud entirely, and can otherwise cause pulse inefficiency and phase errors due to inhomogeneous intensities and laser phases across the cloud. If the position of the atoms is a distance $L$ from the pivot point of the interferometer beam then, to the atoms, the interferometer beam appears translated by an amount $\approx L \theta$ where $\theta$ is the angle of the interferometer beam. For a 10\;m baseline atom interferomter at the latitude of Batavia, Illinois (the location of the MAGIS-100 experiment at Fermi National Accelerator Laboratory) the scale of this translation $L \theta = (L)(\Omega_{\oplus} \cos[\theta_{\text{lat}}])(T) \approx 0.7 \text{ mm}$, which is small relative to the size of a cm-scale interferometry laser beam, and the fact that the atoms see a slightly changed intensity profile during the course of an experiment cycle has a small impact on the measurement outcome. We have defined $\Omega_{\oplus}$ to be the rotation rate of the earth, and $\theta_{\text{lat}}$ to be the latitude of Batavia. Here we have taken $L = 10\text{ m}$, $\Omega_{\oplus} = 7.29\times10^{-5} \text{ rad/s}$, $\theta_{\text{lat}} = 41.85^{\circ} \times\frac{\pi}{180^{\circ}} \text{ rad}$, $T = 1.3 \text{ s}$, and the radial waist of the interferometer beam to be $1\;\text{cm}$ \cite{MAGIS-100:2021etm}. The increased lever arm and longer interrogation times associated with 100\;m baseline interferometry (e.g., $L = 100\text{ m}$, $T = 4 \text{ s}$) increases the scale of this translation to $\approx 2.2 \;\text{cm}$, which is larger than the radial beam waist, so an alternate scheme for rotation compensation is required in order to keep the atom cloud centered on the interferomter beam in the presence of rotation compensation. One solution is to make the size of the beam larger--however, with laser power limited by practical constraints, increasing the beam size lowers the Rabi frequency, which has two detrimental effects: 1. Smaller Rabi frequncies leave atom optics pulse efficiencies more susceptible to the Doppler detuning spread of the atom cloud, and 2. The achievable Rabi frequency can limit the number of pulses that can be driven in a given time, especially for single-photon atom optics on relatively weak clock transitions, which can limit the number of LMT enhancement pulses that can be performed in the interferometer. 

%paragraph outlining the solution
Here, we argue that if one can tune the position about which the inteferometer beam pivots along the axis of the interferometer, and one has control over the initial position and launch angle of the atom cloud, rotation compensation can be achieved in such a way that the atoms stay centered with the center of the interferometer beam during the times that the laser is pulsed on and interacting with the atoms even when the size of the beam is small compared to the translational range spanned by the beam during an experiment cycle. The capability of tuning the initial cloud position and launch angle for individual atom sources is built into MAGIS-100\cite{MAGIS-100:2021etm}.
\begin{comment}
The directionality of the 1D optical lattice which launches the atoms at each connection node will be adjusted via piezo actuated mirrors
\end{comment}

\begin{figure}%[H]%[hbt!]
\begin{centering}
\includegraphics[width=3.37in]{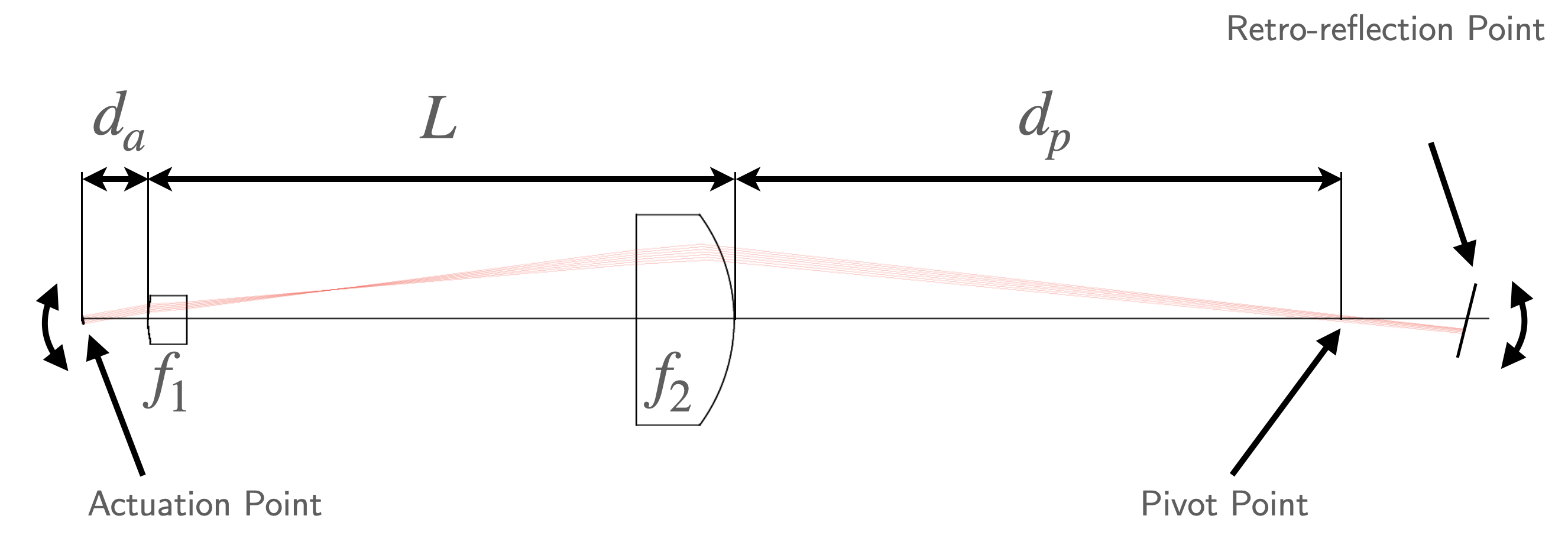}
\caption{\label{fig:pivot_point_optics_cartoon} The optical configuration associated with the pivot point method: By actuating the angle of the interferometer beam a distance $d_a$ from the first telescope lens, the beam pivots about a point that is a distance $d_p$ from the second telescope lens. A subsequent mirror is angled so that the interferometer beam is retro-reflected. The telescope length $L$ is taken to be the sum of the focal lengths which compose the telescope, $L=f_1+f_2$. In this conceptual illustration, the scale of $L$ is on the order of the scale of $d_p$, but in the MAGIS-100 beam delivery system, $d_p$ will in general be one order of magnitude larger than $L$.}
\end{centering}
\end{figure}

The optical configuration associated with this pivot point scheme consists of having piezo-actuated mirrors on each end of the baseline and tuning the path length of the interferometer beam before a telescope to set the position of the pivot point. In MAGIS-100, a telescope will magnify the beam waist to approximately $1\;\text{cm}$ before the beam interacts with the atoms. We define $d_a$ to be the distance between the first telescope lens and the point about which the pointing of the interferometer beam is actuated (the `actuation point'). The position of the pivot point can be adjusted by adjusting $d_a$, as discussed in section \ref{sec:pivot_point_location_tuning}. As indicated in Figure \ref{fig:pivot_point_optics_cartoon}, the `pivot point' is defined as the position after the telescope for which angular adjustments about the `actuation point' manifest as purely angular adjustments in the beam, with no associated translations. A mirror further along the beam's propagation path rotates with the rotating beam so that the retro-reflected beam is colinear with the incident beam.

In Section \ref{sec:dual_isotope_interferometry}, we outline the implemenation of the pivot point scheme for the case of performing interferometery over a $100 \text{ m}$ baseline with a single atom source, as in the case of one of the dark matter search operating modes of MAGIS-100.  In this operating mode, co-located, simultaneous atom interferometers using different isotopes of strontium will span the length of the baseline\cite{MAGIS-100:2021etm}.
%the dual-isotope dark matter search operating mode of MAGIS-100 (using co-located, simultaneous atom interferometers with two isotopes) \cite{MAGIS-100:2021etm}.
In Section \ref{sec:gradiometer_mode}, we extend the scheme to the case of performing atom gradiometry over $100\;\text{m}$ baselines, as in the gravitational wave detection operation mode of MAGIS-100. In Section \ref{sec:longer_baseline_prospects}, we explore the prospects for, and challenges of, scaling this scheme to km-scale baselines.

\begin{figure}%[H]%[hbt!]
\begin{centering}
\includegraphics[width=3.37in]{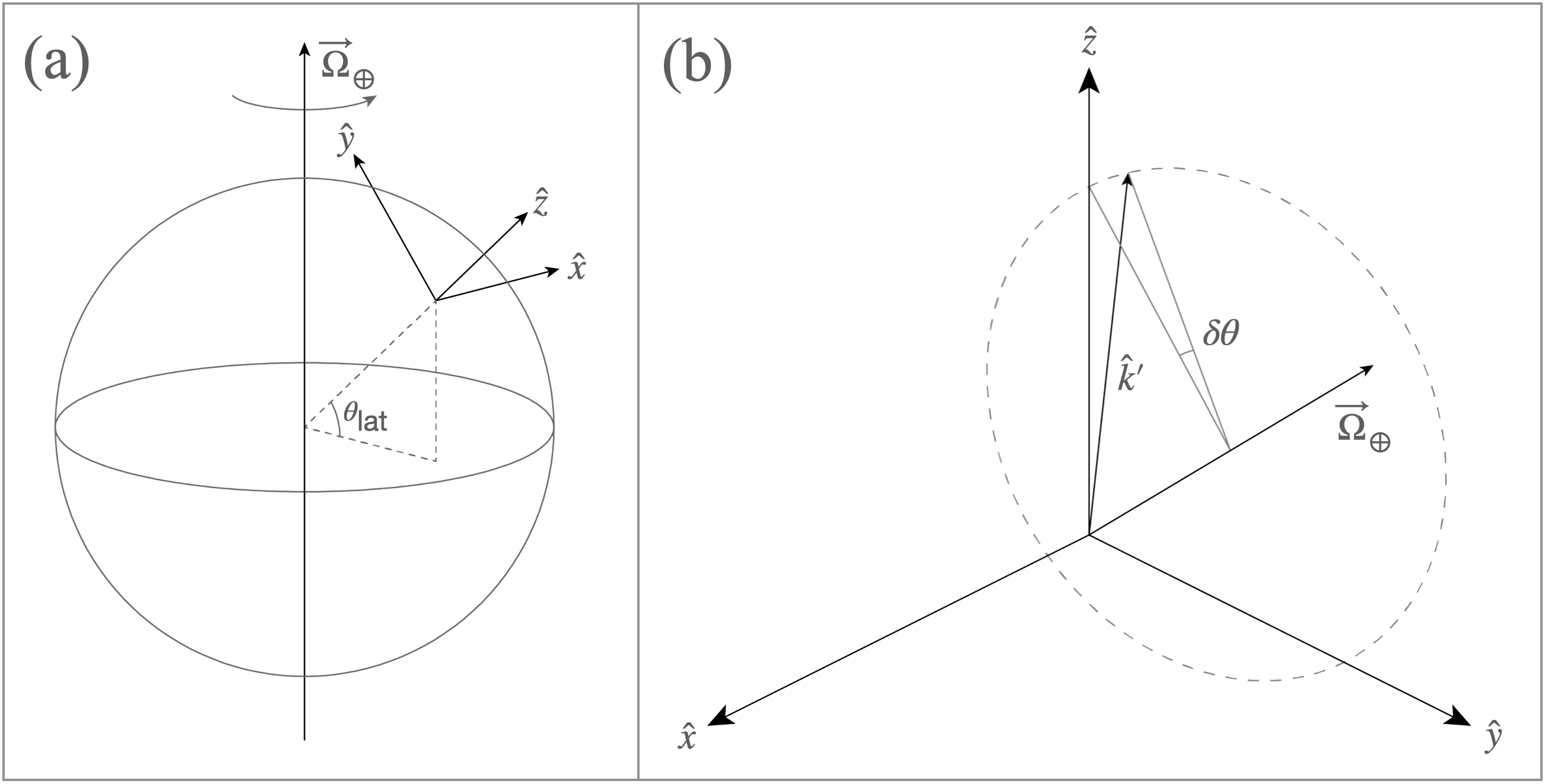}
\caption{\label{fig:rotation_vectors} The relationship between the unit vectors which define the lab frame Cartesian coordinate system $(\hat{x},\hat{y},\hat{z})$ and the rotation vector of the earth $\vec{\Omega}_{\oplus}$ as seen in (a) the space frame, and (b) the lab frame. In the limit that the angles of the interferometer beam $\delta\!\theta$ required to counter-rotate against the rotation of the earth are small, $\delta\!\theta \ll 1$, the required rotations are primarily around the $y$ axis.}
\end{centering}
\end{figure}
%files for this graphic are on  the northwestern desktop -- 2023-9-29

%paragraphic arguing in favor of just considering rotations about the y axis
We argue that to first order in small angular deviations, it suffices to consider rotations about a single axis orthogonal to the vertical direction. Figure \ref{fig:rotation_vectors} shows the relationship between the unit vectors which define the lab frame Cartesian coordinate system $(\hat{x},\hat{y},\hat{z})$ and the rotation vector of the Earth $\vec{\Omega}_{\oplus}$. Since the axis about which the Earth rotates is not orthogonal to the interferometer axis, $\hat{z}$, perfectly rotating the interferometer beam against the rotation of the Earth would in general require rotations about the axis $\vec{\Omega}_{\oplus} = \Omega_y \hat{y} + \Omega_z \hat{z}$ corresponding to the rotation vector of the Earth. For small enough interrogation times $T$, the required rotations of the interferometer beam are primarily around the $y$ axis. Here we choose the coordinate system so that the interferometer axis is along $\hat{z}$, the $y$ axis points toward the north pole, and the $x$ axis points eastward. The components of Earth's rotation vector in these coordinates have magnitudes $\Omega_y = |\vec{\Omega}_{\oplus}| \cos [\theta_{\text{lat}}]$ and $\Omega_z = |\vec{\Omega}_{\oplus}| \sin [\theta_{\text{lat}}]$. Consider a rotation matrix $\hat{R}_{\oplus}[\delta\!\theta]$ which performs rotations by a small angle $\delta\!\theta$ around the vector $\vec{\Omega}_{\oplus}$. To study the scale of the required rotations of the interferometer beam, we take the beam to be initially pointing along the $z$ axis, $\hat{k} = \hat{z}$, and consider the effect of a small Coriolis force compensating rotation on $\hat{k}$, $\hat{k}' = \hat{R}_{\oplus}[\delta\!\theta]\hat{z}$. To first order in $\delta\!\theta$, the required rotation of the interferometer beam is solely about the $y$ axis, in the $x-z$ plane: $\hat{k}' = \hat{R}_y\big[\cos[\theta_{\text{lat}}] \delta\!\theta \big]\hat{z} + \mathcal{O}[\delta\!\theta^2]$, where $\hat{R}_y[\theta]$ is a matrix which rotates a vector an angle $\theta$ about the $y$ axis.
\begin{comment}It is only to capture rotations to second order in $\delta\!\theta$ that we need to consider rotations about the $z$ axis. To second order in $\delta\!\theta$, the full required rotation can be written as a rotation about the $y$ axis followed by a rotation about the $z$ axis, $\hat{k}' =
\hat{R}_z\Big[\frac{1}{2}\sin[\theta_{\text{lat}}] \delta\!\theta\Big]\hat{R}_y\Big[\cos[\theta_{\text{lat}}] \delta\!\theta\Big]\hat{z} + \mathcal{O}[\delta\!\theta^3]$.
\end{comment}

For a single Mach-Zehnder interferometer with interrogation time $T = 4 \text{ s}$, which is indicated in Figure \ref{fig:dual_isotope_mode_rotation_cartoon} and which represents an upper bound on the value of $\delta\!\theta$ which would be required in MAGIS-100, $\delta\!\theta = |\vec{\Omega}_{\oplus}| T \approx 0.29  \text{ mrad}$ with $|\vec{\Omega}_{\oplus}| \approx 7.29\times10^{-5}\frac{\text{rad}}{\text{s}} $. At this maximum angular range, the angle that is made by the $z$ axis and the projection of $\hat{k}'$ onto the $x-z$ plane is $\cos[\theta_{\text{lat}}]\delta\!\theta + \mathcal{O}[\delta\!\theta^2] \approx 0.22 \text{ mrad}$, while the angle that is made by the $z$ axis and the projection of $\hat{k}'$ onto the  $z-y$ plane is $\frac{1}{2} \sin[\theta_{\text{lat}}]\cos[\theta_{\text{lat}}] \delta\!\theta^2 +\mathcal{O}[\delta\!\theta^3] \approx 21 \text{ nrad}$.  For simplicity of illustration in figures, we will ignore the small projection of $\hat{k}'$ onto the z-y plane, though the rotation compensation system will have the capability to implement the corresponding small corrections to the angle of $\hat{k}'$ as needed.

In considering the overlap between the interferometry laser beam and the atoms, it is also important to consider deflections of the atom trajectories due to Coriolis forces. The predominant deflections in the trajectories are in the $x-z$ plane because the Coriolis force couples the motion of an atom along the $x$ axis with the large `upward' component of its launch velocity, $v_z$, and does not couple the $y$ component of the motion to $v_z$ since the axis of rotation is in the $y-z$ plane. We estimate residual deflections along the $y$ axis for the scheme indicated in Figure \ref{fig:dual_isotope_mode_rotation_cartoon} to be $\approx 4.5 \;\mu\text{m}$. To simplify the illustration of the pivot point method, we do not consider here dynamics of the atom trajectory in the $y$ direction and focus on the $x-z$ plane. We note also that the centrifugal force produces a constant acceleration in the $y-z$ plane which can be thought of as contributing a small correction to a pure gravitational acceleration: $\vec{g}_{\text{eff}} = (g - R_{e} \Omega_y^2)\hat{z} + (R_{e} \Omega_y \Omega_z)\hat{y}$, where $R_e$ is the radius of the Earth (taking the Earth to be a perfect sphere)\cite{morin_2008}. In practice, the interferometer beam will be aligned to the direction in which the atoms freely fall, which is along $\vec{g}_{\text{eff}}$.  Taking this effect into account, the $\approx 4.5 \;\mu\text{m}$ deflection cited above can be understood as referring to a deflection along a $y'$ axis, where the primed coordinates refer to a coordinate system which has been slightly rotated around the $x$ axis so that the $z'$ axis points along $\vec{g}_{\text{eff}}$.
%this math is worked out in 2023-9-18 (pivot point paper)

\begin{comment}
The new direction the interferometer beam point is given for small angels $\delta\!\theta$ by $\hat{k}' = (\cos[\theta_{\text{lat}}]\delta\!\theta)\hat{x} + (\frac{1}{2}\cos[\theta_{\text{lat}}]\sin[\theta_{\text{lat}}]\delta\!\theta^2)\hat{y} + (1-\frac{1}{2}\cos[\theta_{\text{lat}}]^2\delta\!\theta^2)\hat{z} + \mathcal{O}[\delta\!\theta^3]$. To first order in $\delta\!\theta$, the required rotation of the interferometer beam is solely in the $x-z$ plane, rotating at a rate of $\Omega_y = |\vec{\Omega}_{\oplus}|\cos[\theta_{\text{lat}}]$. (SIGN CORRECT ON $\delta\!\theta$?). To first order in the required rotation angle $\delta\!\theta$, in order to counter rotate the interferoemter beams with the rotation of the earth for interferetomer sequences with $4 \text{ s}$ interrogation times, the interferometer beam will be tilted by an angle about the $y$ axis of $\Omega_y T \approx $

for 

the second order corrections in $\delta$

$|\vec{\Omega}_{\oplus}| T \approx (7.27\times10^{-5}\frac{\text{rad}}{\text{s}})(4 \text{s}) \approx 0.29  \text{mrad}$. The second order corrections require an additional tilt about the $y$ axis of 
\end{comment}

\subsection{Coriolis Force Compensation for Long Baseline Interferometry with a Single Atom Source}\label{sec:dual_isotope_interferometry}
\begin{figure}%[H]%[hbt!]
\begin{centering}
\includegraphics[width=3.37in]{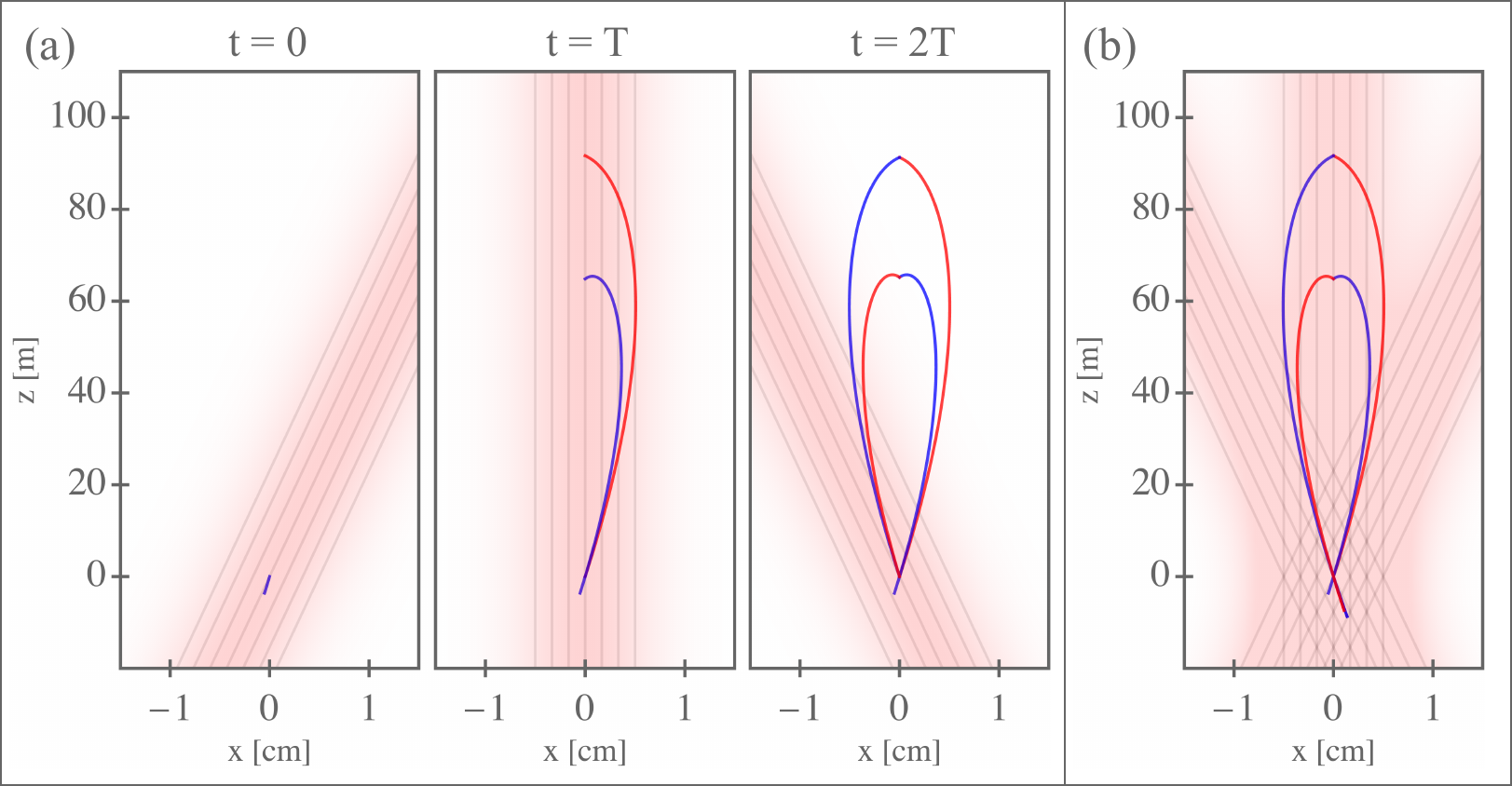}
\caption{\label{fig:dual_isotope_mode_rotation_cartoon} A diagram demonstrating the basic pivot point idea for a Mach-Zehnder sequence. (a) The angle of the interferometer beam at the time of the beamsplitter pulse $t=0$, the mirror pulse $t=T$, and the final beamsplitter pulse $t=2T$, along with the paths traced by the classical trajectories of the two arms of the interferometer at each of those three times. (b) The total classical paths traversed by the upper and lower interferometer arms along with each of the three angles of the interferometer beam. For the dual-isotope operating mode of MAGIS-100, the momentum kicks will be achieved via two-photon Bragg transitions on the 679\;nm resonance of strontium \cite{MAGIS-100:2021etm}. In this case, the blue line denotes the classical path traces by the part of the atom wavefunction in the $|0\hbar k \rangle$ state which has not been kicked by any photons, and the red line denotes path traces by the atom in the $|1000 \hbar k \rangle$ state that has been kicked by 1000 photons. The radial beam waist of the beam is 1\;cm, and the rays are separated by a distance of 1/6\;cm.}
\end{centering}
\end{figure}

%paragraph describing the dual isotope interferometry figure
In Figure \ref{fig:dual_isotope_mode_rotation_cartoon}, a Mach-Zehnder pulse sequence with a $T = \;4 \text{ s}$ spacing between beam splitter and mirror pulses and LMT number $n=1000$ is performed, wherein an atom cloud is launched at an angle of $\approx 0.14 \text{ mrad}$ from the $z$ axis with a total velocity of $\approx$ 35.86 m/s. The launch angle is chosen so that the two arms of the interferometer have transverse position $x = 0$ at the time of the mirror pulse, and so that the position of the atom cloud is the same for both the first and final beam splitter times.
%and so that the positions of both interferometers arms are the same at the first and final beamsplitter times.
%and so that the position of each arm is the same at the final beamsplitter pulse as it was at the first beamsplitter pulse.
The radius of the earth is taken to be $R_e$ = 6378 km.
%(which slightly effects the freefall of hte atoms due to the centrifuglar force).
At time t = 0, an LMT beamsplitter pulse is performed by an interferometer beam angled by $\Omega_y T \approx 0.22 \text{ mrad}$ about the origin of the lab frame coordinate system which imparts a momentum kick on the upper interferometer arm of $ 1000 \hbar k$ (for simplicitly, we assume that the momentum kicks happen instantaneously), where $k$ is the wave number of the 679\;nm light that is resonant with the $ ^3P_0 \leftrightarrow \!\!\! \ ^3 S_1 $ resonance of strontium (for the dual-isotope operating mode of MAGIS-100, atom optics will be performed via two-photon Bragg transitions on the 679\;nm resonance of strontium \cite{MAGIS-100:2021etm}). The atoms then propagate in a superposition of different momentum states subject to Coriolis and centrifugal forces before the mirror pulse at time $ t=T $ imparts a $1000 \hbar k$ momentum kick downward on the upper interferometer arm and a kick with equal magnitude and opposite direction on the lower arm. The atoms then freely propagate for a time $T$ before spatially recombining, at which point a final beamsplitter pulse is performed by the beam angled symmetrically about the $z$ axis with respect to its angle at time $t = 0$. The choice of pivot point and initial atom cloud kinematics allows for the beam to stay centered on the atom cloud for each interaction time while the interferometer beam is rotated to suppress Coriolis dependent phase shifts and contrast loss.  In this example, the pivot point is located at the initial position of the atoms at $t=0$.
%this math is written out in 2023-9-29 (coriolis force).nb
%then later, if we have an initial draft, time permitting -- we can put in some of the math of how the launch angles are determined. Could be nice, but not critical.

\begin{comment}
The counter rotation of the interferometer beam provides a suppression of the phase shift on the initial kinematics of the atom, which under an ensemble average of the thermal atom cloud results in interferometer contrast loss
\end{comment}

%note: this figure is saved in 2023-9-29 on northwestern desktop

\subsection{\label{sec:pivot_point_location_tuning}Tuning the Location of the Pivot Point}

\begin{comment}
\begin{figure}[H]%[hbt!]
\begin{centering}
\includegraphics[width=3.0in]{pivot_point_method/pivot_point_as_a_function_of_position_diagram.png}
\caption{\label{fig:pivot_point_diagram_cartoon} placeholder for diagram indicating meaning of $d_m$ and $d_p$.}
\end{centering}
\end{figure}
\end{comment}
\begin{comment}
%starting with example from 
%https://tex.stackexchange.com/questions/55161/how-to-arrange-image-and-text-to-appear-side-by-side
\begin{wrapfigure}{r}{0.82in}
\begin{center}
\includegraphics[width=0.82in]{pivot_point_method/zemax_dp_dm_cartoon_draft_1.png}
%\caption{\label{fig:pivot_parameter_space} test caption}
\end{center}
\end{wrapfigure}
\end{comment}

The location of the pivot point can be tuned by adjusting the distance between the actuation point and the first telescope lens, $d_a$ (see Fig. \ref{fig:pivot_point_optics_cartoon}). Here we perform ABCD matrix ray tracing calculations in the paraxial and thin lens limits in order to provide an approximate analytic scaling of the pivot point location as a function of $d_a$. We consider physical optics effects, including also the effects of the exact lens geometries, with numerical calculations using the Zemax software package in section \ref{spherical_lens_effects}. Here we consider an optical system consisting of two telescope lenses with focal lengths $f_1$ and $f_2$ separated by a distance $f_1 + f_2$. A ray pivots a distance $d_a$ away from the first telescope lens, and we define $d_p$ to be the distance from the second telescope lens to the pivot point. The relationship between $d_p$ and $d_a$ can be expressed as

\begin{equation}\label{dp_da_equation}
    d_p = (f_2 + M^2 f_1) - M^2 d_a
\end{equation}

\noindent where $ M = f_2/f_1 $ is the magnification of the telescope and the other parameters are indicated in Fig. \ref{fig:pivot_point_optics_cartoon}. For MAGIS-100, two ultrahigh quality spherical lenses (Optimax) \cite{MAGIS-100:2021etm} with focal lengths $f_1 = 150\text{ mm}$ and $f_2 = 4500\text{ mm}$ will be used to magnify a beam focused to a $\approx300\;\mu\text{m}$ radial waist at the position of the tip tilt mirror to a waist of $\approx 1 \text{ cm}$. A vacuum-compatible, piezo-actuated tip-tilt mirror mount has been procured from Mad City Labs (a custom vacuum-compatible version of the Nano-MTA2) which will enable tilting the interferometer beam before this telescope. This mount has a total angular range of $10\text{ mrad}$, and for tilts about $\hat{y}$--which will be made to be approximately orthogonal to the nominal plane of incidence--the total range with which the beam can be angled is $20 \text{ mrad}$ because the change in the angle of an incident beam upon reflection from the actuatable tip-tilt mirror is twice the angle of incidence. The $M=30$ telescope will magnify the size of the interferometer beam by a factor of $M$ and de-magnify the angles produced by this tilted mirror by that same factor, so that the largest angular tilt allowed by this system in one direction is $ \approx 10\text{ mrad}/M \approx 0.33 \text{ mrad}$.  This angular range is large enough to perform the $T=4\text{ s}$ sequence shown in Figure \ref{fig:dual_isotope_mode_rotation_cartoon}, whose maximum angular range is $\approx 0.22 \text{ mrad} $ in one direction. The tip-tilt retro-reflection mirror\cite{MAGIS-100:2021etm} positioned $\approx 100\text{m}$ below this second telescope lens will be angled to keep the reflected beam aligned with the downgoing beam.

\subsection{\label{sec:tip_tilt_system_mechanical_design} Tip-Tilt System Mechanical Design}

\begin{figure}%[H]%[hbt!]
\begin{centering}
\includegraphics[width=3.37in]{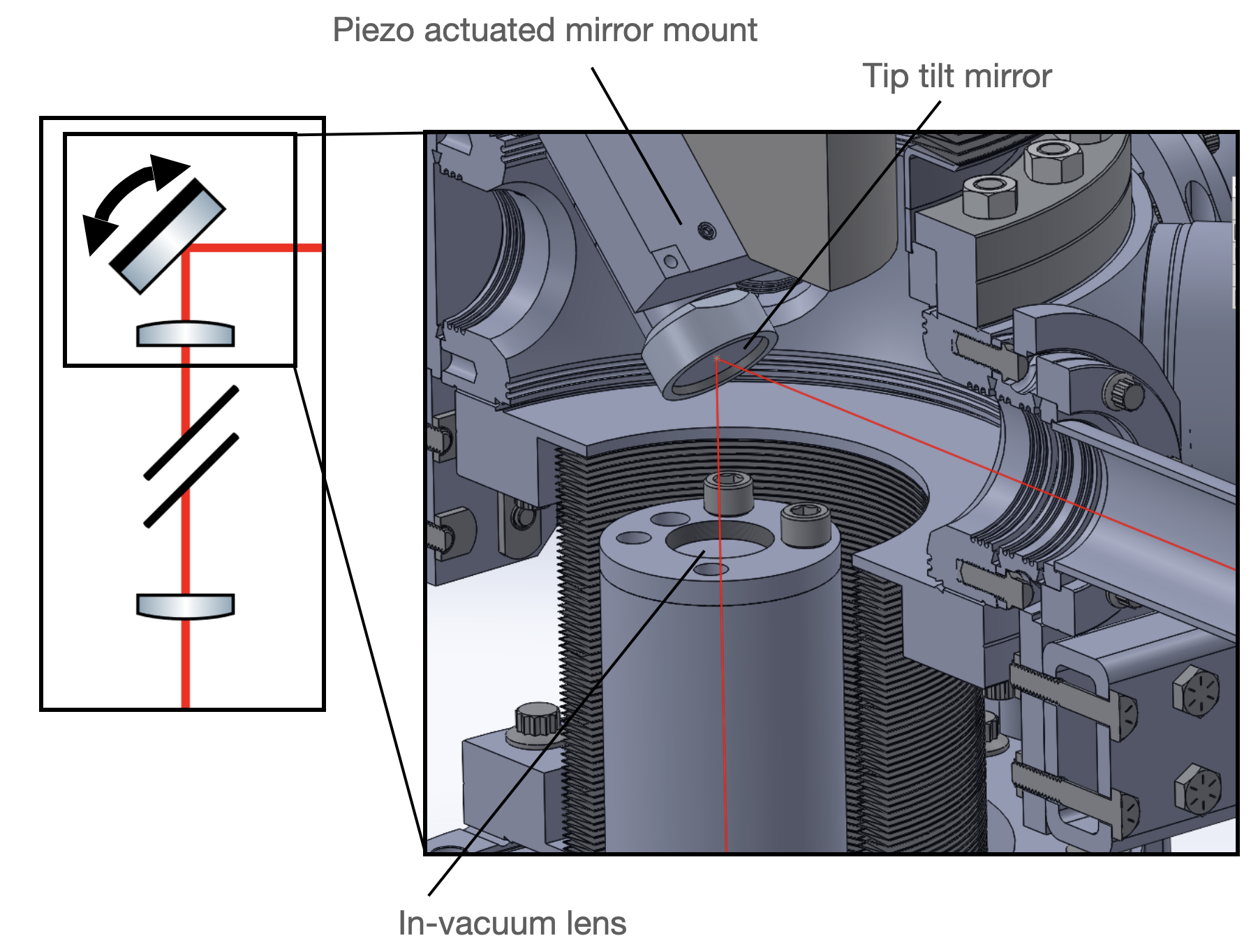}
\caption{\label{fig:pivot_point_optics_CAD} CAD model of the in-vacuum tip-tilt mirror and first telescope lens.}
\end{centering}
\end{figure}

As shown in Figure \ref{fig:pivot_point_optics_CAD}, our design of this section of the tip-tilt system involves the piezo-actuated stage mounted at a $45^{\circ}$ angle on a steel post. Out of vacuum stepper motors will adjust the distance between the tip tilt mirror and the first telescope lens to set the pivot point, and the piezos in the Mad City Labs mount will rotate the mirror during an experiment cycle for Coriolis force compensation. We will have the capability to translate the beam with out-of-vacuum stepper motors, which could be useful for fine tuning the alignment of this beam.

%this is the section on the lever arm, which we found wasn't so relevant.
\begin{comment}
Since the pivot point of the mirror is not exactly at the surface of where the interferometer beam hits the mirror, adjustments in the angle of the mount will induce slight translations in the tip tilt mirror. The lever arm between the axis about which the mount rotates and the point at which the interferometer beam reflects off the tip tilt mirror is $L \approx 19 \text{mm}$, and the translation induce as a function of the angle of this mount, assuming perfect alignment at an angle of $\delta\!\theta = 0$ is $\Delta x = \frac{L \delta \! \theta^2}{\sqrt{2}} + \mathcal{O}[\delta\!\theta^3] \approx 1.3 \mu \text{m}$ for the full $10\text{ mrad}$ angle. (DOUBLE CHECK THIS MATH WITH PROF KOVACHY -- BEN SAYS THE AXES OF ROTATION REFERENCE THE CENTER OF THE CYLINDRICAL PIECE). This translation in the interfereomter beam is magnified by the $M=30$ telescope. Out of vacuum optics mounted to stepper motors actuated stages will compensate for any residual translations during an experiment cycle, as well as allow for more dynamic control over the translation of the beam more generally (see Figure \ref{fig:shaft_side_optics_cartoon}. (IN THE CASE OF DOING LMT AND SHIFTING THE BEAM BY EACH PULSE).(MAYBE THORLABS LNR25ZFS) (IS THIS EVEN NECISSARY?)
\end{comment}

\subsection{\label{sec:spherical_lens_aberrations}Effect of Spherical Lens Aberrations} \label{spherical_lens_effects}

In this section, we determine the scale of the aberrations associated with the interplay of the pivot point system and the spherical nature of the telescope lenses through physical optics simulations using the Zemax software package, which capture non-paraxial effects along with effects from the finite extent of the curved face of the telescope lenses. We study the aberrations induced by the spherical telescope lenses on the intensity and phase profiles of the interferometer beam under $0$ and $10\text{ mrad}$ beam tilts at five different propagation distances between $0$ and $200 \text{ m}$ after the second telescope lens. We find that the scale of the aberrations induced on the interferometer beam by the spherical nature of the telescope lenses are comparable to the expected phase deviations from surface imperfections in the optics themselves. 

Spherical lenses are cheaper to manufacture but induce some anharmonicity into the phase profile of a transmitted Gaussian beam. The dominant mechanism by which aberrations in the interferometer beam are expected to limit the sensitivity of the MAGIS-100 interferometer is the coupling of aberrations to laser pointing jitter and to shot-to-shot fluctuations in initial atom cloud kinematics \cite{MAGIS-100:2021etm}. 

%maybe some room for a calculation here, if time

\begin{figure}[H]%[hbt!]
\begin{centering}
\includegraphics[width=3.37in]{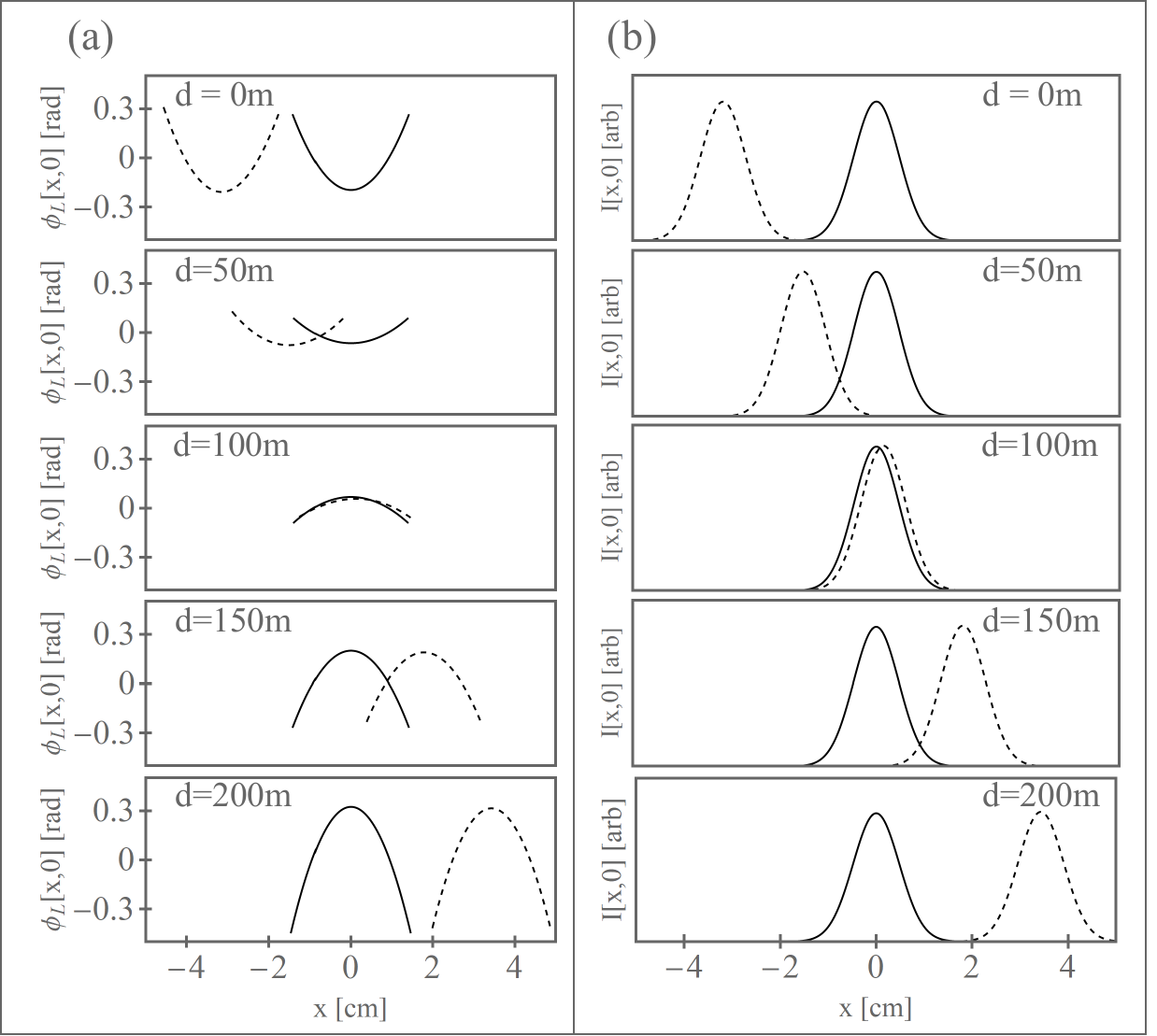}
\caption{\label{fig:untilted_and_tilted_together} (a) Cross sections of interferometer beam (a) phase profiles and (b) intensity profiles after propagating through the spherical telescope lenses with no initial tilt (solid line) and with a $10\text{ mrad}$ tilt $5\text{ cm}$ before the first telescope lens (dotted line) for a few different propagation distances $d$ after the second telescope lens. Here the linear component of the tilted beam's phase profile, computed via geometric optics, is ignored. $\Delta\phi_L[x,y]$ is defined to be the phase profile of the interferometer beam, where $x$ and $y$ are the dimensions transverse to the propagation axis of the beam, and $I[x,y]$ is defined to be the intensity profile of the beam}
%the contribution of a linear component of the tilted beam's phase profile (the linear component is calculated via geometric optics) is ignored.}

\begin{comment}
(b) The difference in the phase profiles at different propagation distances $d$. (d) cross sections of the intensity profile of the interferometer beam after propagation through the spherical telescope lenses with no tilt (solid line) and a tilt of $10\text{ mrad}$ before the first telescope lens. (c) the relative difference in the intensity profiles between the tilted and untitled cases for different propagation distances $'d'$ beyond the second telescope lens.
\end{comment}

\end{centering}
\end{figure}

\begin{figure}[H]%[hbt!]
\begin{centering}
\includegraphics[width=3.37in]{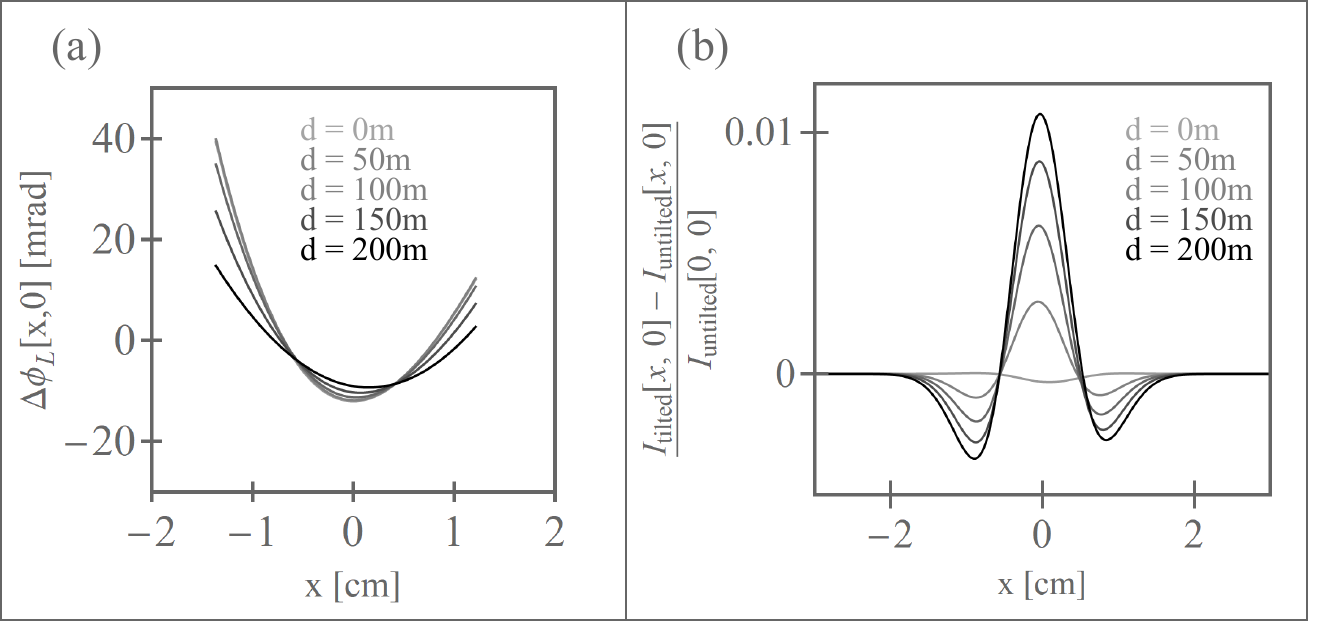}
\caption{\label{fig:difference_in_untilted_and_tilted} (a)  The difference in the phase profiles and (b) the relative difference in intensity profiles at different propagation distances $d$ after the second telescope lens between the untilted interterferometer beam and a beam tilted by $10\text{ mrad}$ a distance of $5 \text{ cm}$ before the first telescope lens. Here the difference is taken after the tilted interferometer beam is shifted in $x$ at each propagation distance $d$ by the expected position of the beam along $x$ as determined by geometric optics.}
\end{centering}
\end{figure}

To investigate aberrations associated with the interplay of the tip/tilt system and the telescope, we start with a Gaussian beam profile at a focus with a radial waist $300\;\mu\text{m}$ incident on the tip/tilt mirror. The first lens has a plano-convex shape and is oriented with its curved face pointing away from the center of the telescope a distance $50\text{mm}$ from the tip-tilt mirror. The radius of curvature of this lens is $R_1 = 68.295 \text{mm}$, and its center thickness is $t_1 = 3\text{ mm}$. The second telescope lens is also plano-convex and also has its curved face pointing away from the center of the telescope. This lens has radius of curvature $R_2 = 2048.85 \text{mm}$ and a center thickness of $ t_2 = 7.5\text{ mm}$. The distance between the curved faces of the lenses is equal to the sum of their focal lengths $L=f_1+f_2 = 4650\text{ mm}$, where $f=\frac{R}{n-1}$. Both lenses are made of a material whose index of refraction is approximately $n = 1.455$ (fused silica). The parameter $d$ in Figures \ref{fig:untilted_and_tilted_together} and \ref{fig:difference_in_untilted_and_tilted} denotes the distance from the curved face of the second telescope lens to the plane in which the intensity and phase profiles are evaluated. Here we plot only the cross section of the phase and intensity profiles along the $x$ axis for different distances $d$ along the $z$ axis, which for rotations about the $y$ axis is where the dominant sources of aberration are under rotations of the beam.

%these figures were generated in 2023_9-29.nb on the northwestern desktop then modified in adobe illustrator

For ease of visualization, the phase profile plots ignore a phase factor varying linearly in $x$ that corresponds to the angle of the beam as calculated purely by ray tracing. We see no remarkable aberrations in the intensity profile of the beam, and the aberrations to the phase profile, as seen in Figure \ref{fig:difference_in_untilted_and_tilted}, are on length scales on the order of the beam size.  Over a $\pm 9\text{ mm}$ range centered on the center of the beam, the phase aberrations have a maximum magnitude of $\approx 12 \text{ mrad}$, which is comparable to the expected phase deviations from surface imperfections in the optics themselves. The size of the atom cloud is $\approx 3 \text{ mm}$ or less, so beyond this range, most atoms will not sample the inhomogeneity in the laser phase profile.
%for the 12mrad number, check 2023-8-24.nb on northwestern computer

\subsection{\label{sec:gradiometer_mode} Coriolis Compensation for Atom Gradiometry}

\begin{figure}%[H]%[hbt!]
\begin{centering}
\includegraphics[width=3.37in]{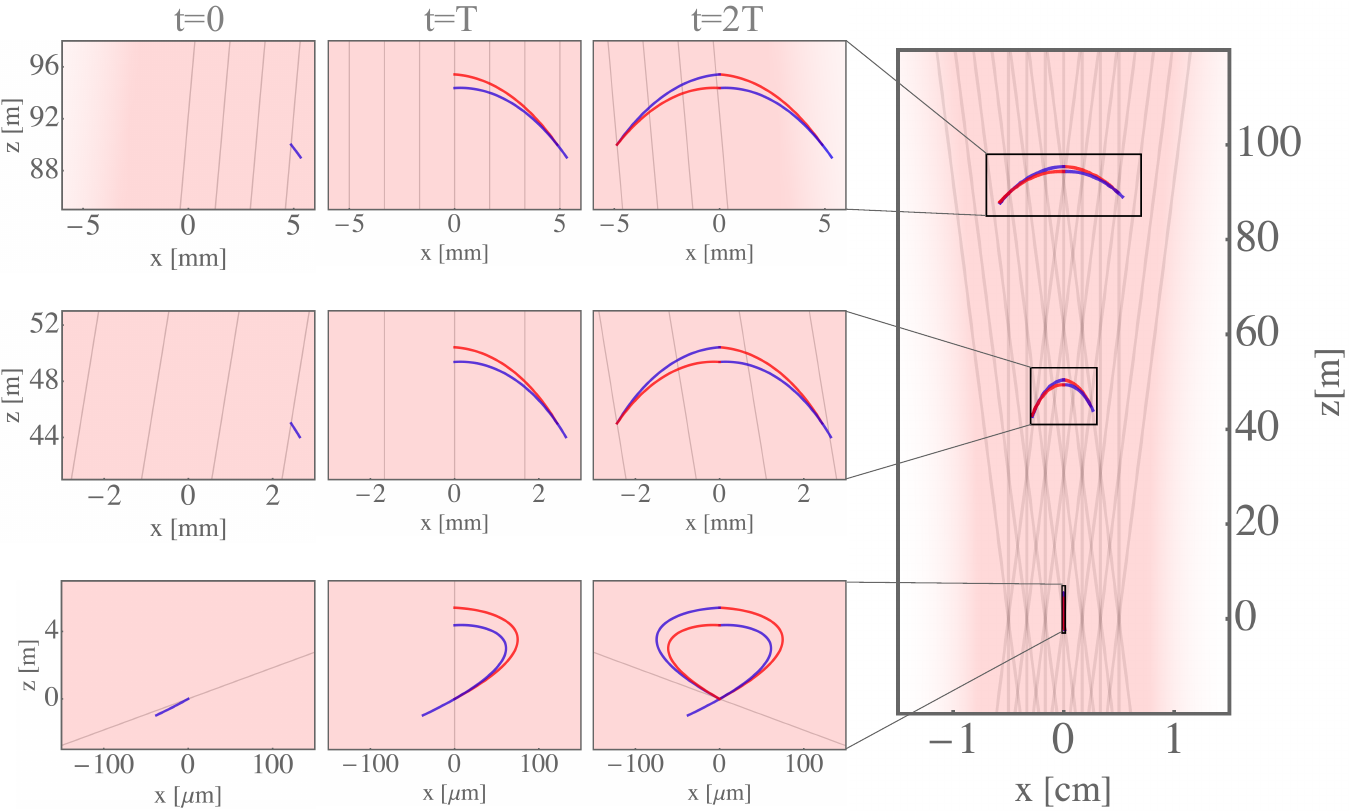}
\caption{\label{fig:gradiometer_mode} Illustration of rotation compensation with the pivot point method in gradiometer mode.  Three atom interferoemters are initially located at respective vertical positions of  $z_0 = 0 \text{ m}, 45 \text{ m}, 90\text{ m}$. The spacing between beam splitter and mirror pulses if $T = 1$\;s, and 160$\hbar k$ LMT is used. Note that the horizontal axis for each interferometer has a different range. In this example, the pivot point is chosen to be at the location of the bottom atom source as in Figure \ref{fig:dual_isotope_mode_rotation_cartoon}, but the pivot point could instead be located at the middle atom source to reduce the size of the required initial position offset of the topmost atom cloud.}
\end{centering}
\end{figure}

An atom gradiometer is an apparatus in which multiple atom interferometers are spaced over a long baseline. MAGIS-100 is designed to be able to operate in this mode for mid-band gravitational wave detection and certain dark matter searches \cite{MAGIS-100:2021etm}. The influence of Coriolis forces can be suppressed in a gradiometer configuration by performing multiloop sequences \cite{Dubetsky_2006, MAGIS-100:2021etm}, but this approach also suppresses the response of the interferometer to low frequency signals. The pivot point method outlined in Section \ref{sec:outline_of_pivot_point_method} 
%Coriolis force compensation over long baselines by rotating the interferometer beam
could be used in a gradiometer mode where the signals of interest are of a lower frequency. Performing long baseline rotation compensation with the pivot point method 
%The case of performing 'pivot point' rotation compensation
for atom gradiometry relies on the ability to adjust the initial positions and velocities of the different atom clouds individually. In Figure \ref{fig:gradiometer_mode}, three atom interferometers are separated by 45\;m and span roughly the length of the 100 m baseline. The sequence presented in Figure \ref{fig:gradiometer_mode} is a Mach-Zehnder interferometer with a $T=1\text{ s}$ interrogation time and $160\hbar k$ momentum transfer at each atom-laser interaction time. The atom clouds from bottom to top are launched with respective angles of approximately $35.09 \;\mu\text{rad}$, $-0.23 \text{ mrad}$, and $-0.49\text{ mrad}$ relative to the $z$ axis, each with a total initial velocity of $\approx 9.26 \text{ m/s}$. The interferometer beam is initially angled by $\Omega_y T \approx 54.3 \; \mu\text{rad}$, and the atom clouds are prepared with a lateral offset from the interferometer baseline of $\approx 0, 2.44\text{ mm}, \text{and } 4.87 \text{ mm}$, read from bottom to top.

%the rendering of the plots was done in 2023-9-14 (coriolis force) on personal laptop, then images we organized in adobe illustrator on northwestern desktop

\subsection{\label{sec:longer_baseline_prospects}Prospects For, and Challenges of, Scaling Up to Longer Baselines}

\begin{figure}%[H]%[hbt!]
\begin{centering}
\includegraphics[width=2.5in]{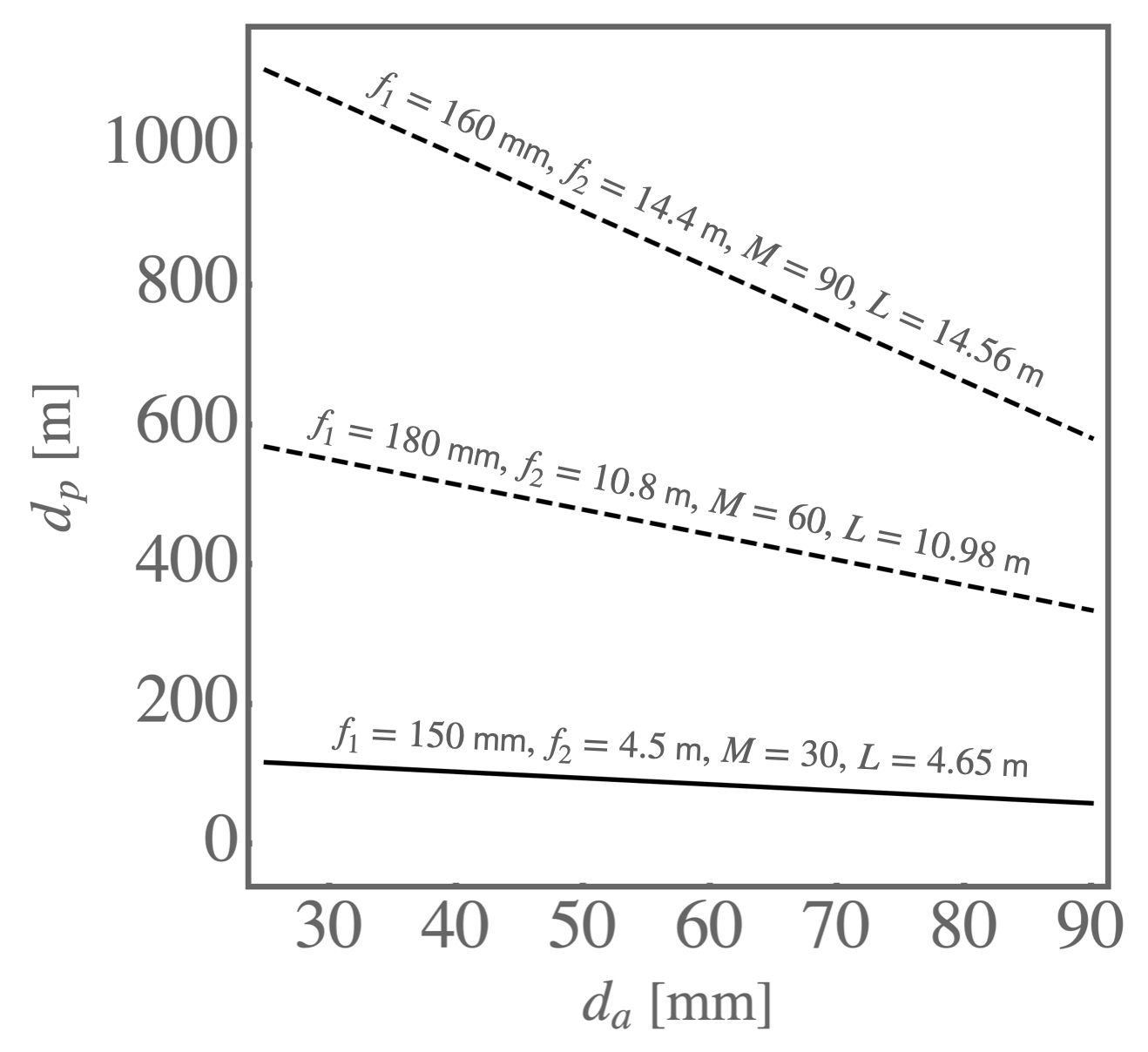}
\caption{\label{fig:pivot_point_parameter_space} Possible $d_p[d_a]$ for different choices of $f_1$ and $f_2$, referencing Equation (\ref{dp_da_equation}). The dependence of $d_p$ on $d_a$ for three different lens combinations are considered here in order to illustrate some possible lens combinations for interfereomters with baselines up to the km scale. The length of the telescope $L$ is defined to be the sum of the focal lengths of the two lenses which compose the telescope, $L = f_1 + f_2$. The solid line corresponds to the MAGIS-100 telescope parameters, and the dotted lines are potential lens combinations for longer baseline interferometers.}
\end{centering}
\end{figure}

The pivot point method can be extended to future larger baseline interferometers, but alternate telescope parameters are required in order to place the pivot point near the end of the longer baseline and ensure the ability to tune it over a range about that point. Only two-lens telescopes are considered here, though telescopes with three or more lenses may also be feasible. Figure \ref{fig:pivot_point_parameter_space} presents a few possible lens combinations, referencing Equation (\ref{dp_da_equation}). If the range over which $d_a$ is adjusted is held fixed as the baseline increases, a larger telescope magnification is required in order to scan the pivot point over a larger range. In general, as one increases the baseline of the interferometer, a longer two-lens telescope is required. One challenge associated with performing Coriolis force compensation with the pivot point method over longer baselines is that the required deflections of the interferometer beam are larger, which requires vacuum pipes with larger internal diameters. The diameter of the second telescope lens also increases with increasing baseline length. For a Mach-Zehnder interferometer sequence wherein the classical trajectories of the atoms span km-scale baselines, the required rotations about the pivot point are roughly $\Omega_y T L \approx 0.7 \text{ m}$ for $L = 1 \text{ km}, \; T = 13 \text{ s}$, and $\Omega_y \approx 54.3 \; \mu \text{rad}/\text{s}$. In order for the interferometer beam to not clip on the edge of optics or on the internal diameter of vacuum pipes, the bottom telescope lens and the internal diameter of the vacuum pipes near the second telescope lens would need to have diameters on the meter scale, which would have significant associated costs.

% note: this image was generated in 2023-7-5 (pivot point paper)
% maybe we could try out other combinations here

\section{\label{sec:beam_delivery_system}Interferometer Laser Beam Delivery System}

\subsection{\label{sec:beam_delivery_system_mechanical_design} Beam Delivery System Mechanical Design}

The laser transport system transports the interferometer beam from where it is generated in the laser room over to the shaft where the interferometry is being done. It uses a combination of stable mechanical mounts, a relay imaging system, and a short fiber to provide both passive pointing stability and a stable multi-Watt interferometer beam power.

Building a dedicated laser room serves two critical functions: temperature control and laser safety. Having a dedicated temperature-controlled room suppresses associated drifts and reduces the frequency of various laser feedback loops falling out of lock. Also, consolidating the laser system in a central, interlocked room is required for laser safety purposes. Constructing a laser room within a facility not purpose-built for atom interferometry presents a challenge: It can be difficult to position the laser room in close proximity to the main experiment without resorting to costly and time-consuming structural modifications to the existing facility in order to accommodate the laser room's spatial footprint. Consequently, this constraint may necessitate situating the laser room at a considerable distance from the main experiment.
% One associated challenge is that laser rooms typically have a substantial spatial footprint, and constructing a laser room in a building not purpose-built for atom interferometry, without expensive and time consuming structural modifications to the building, can impose the requirement that the laser room be far from the atom interferometer.
MAGIS-100 is being constructed in the MINOS Service Building at Fermilab, which was originally built as an access building for underground experiments. Facilities constraints require the interferometry laser beam to be transported to the shaft over $\approx 10 \text{ m}$ horizontally, and also for the beam to be transported at a height of $\approx 3 \text{ m}$ above the floor of the building.

\begin{comment}
%paragraph on the laser transport system
The laser transport system emerges from geometry and facilities constraints of the MINOS building in which MAGIS-100 is being built -- these constraints require the laser lab to be $\approx 10 \text{m}$ away from the shaft and also for the beam to be transported at a height of $\approx 3 \text{m}$ above the floor of the building. 
%this is ~10 ft in 'm'
\end{comment}
A limitation of fiber-based laser transport
%An additional constraint which affects our ability to transfer the beam light through a fiber
is that for multi-Watt laser powers and $\approx 10 \text{m}$ long fibers, losses from stimulated Brillouin scattering become a significant concern\cite{Kobyakov:10}. It is productive to maximize the power of the beam that is delivered to the atoms because larger beam powers result in larger Rabi frequencies, which are advantageous for the reasons outlined in Section \ref{sec:outline_of_pivot_point_method}. We therefore opt to transport the beam in free space. Free space transport, however, poses its own challenges since it is more susceptible to alignment drifts in the delivery optics.

\begin{comment}
    We opted for transporting the beam above ground since it was determined to be less expensive and it (would keep us more isolated from pointing instability from people walking over it? Is it better than having 3m tall towers?). Early vibration analysis performed in ANSYS, along with seismometer data collected in situ \cite{Mitchell_2022} indicated that the required $30 \text{ nrad}/\sqrt{\text{Hz}}$ pointing stability of the interferometer beam \cite{MAGIS-100:2021etm} put too unrealistic a onus on the delivery structures to be vibrationally insensitive, so we opted to have the interferometer beam couple into a structurally stable fiber after propagating through the laser transport system.
\end{comment}

To address these challenges, we include a $\approx 1 \text{ m}$ long, single-mode fiber (short enough that we do not observe stimulated Brillouin scattering) after the free space laser transport to remove the influence of pointing fluctuations from all preceding optics and free-standing support structures. We will then anchor all subsequent optics to a structurally stable floor and wall by the shaft.
%MOVED TO LATER -- The fiber is single-mode and also provides initial spatial filtering of the beam \cite{MAGIS-100:2021etm}.
With this short fiber, pointing fluctuations in the beam manifest as fluctuating fiber coupling efficiencies. After the short fiber, we will pick off a small amount of light to measure the post-fiber power of the beam and feed that signal back to the laser room, where an acousto-optic modulator will adjust and stabilize this power. The lock point must be set slightly below the maximum achievable laser power in order to correct for dips in the post-fiber power of the beam caused by drifts in the laser transport. Improved stability of the laser transport system suppresses the magnitude of these power drops, allowing for a higher lock point, and an increase in the power delivered to the atoms. This combination of a short fiber and a power stabilizing feedback system enables the beam to have both stable pointing and power.
%monitor the power of the post-fiber beam and provide feedback to stabilize the post-fiber power with an acousto-optic modulator in the laser room, enabling the post-fiber beam to have both stable pointing and power. We want the fiber coupling efficiency to be as passively stable as possible because the scale of the post-fiber power fluctuations limits the maximum post-fiber power. This is because greater fluctuations require us to have the `set point' of the laser power feedback system be smaller so that we have enough range to correct for reductions in the fiber coupling efficiency. This power reduction manifests as a smaller atom optics Rabi frequency, which is unfavorable for the reasons outlined in Section \ref{sec:outline_of_pivot_point_method}. 

\begin{figure}%[H]%[hbt!]
\begin{centering}
\includegraphics[width=3.37in]{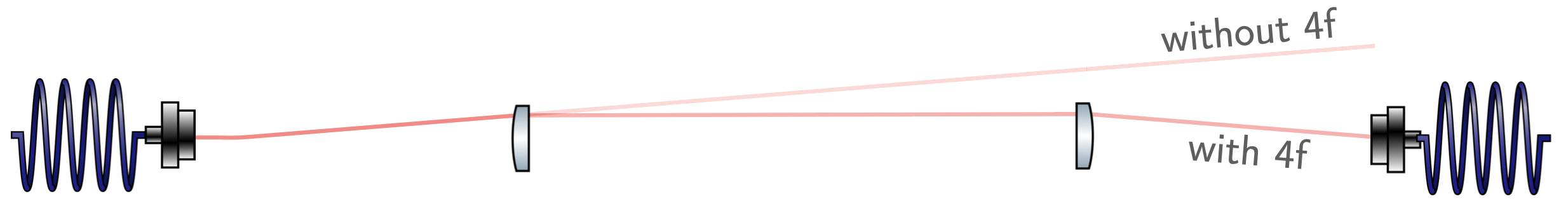}
\caption{\label{fig:relay_imaging_cartoon} The idea behind relay imaging: Lenses in the beam path reduce the susceptibility of the fiber coupling efficiency to drift in optics. Without the relay imaging lenses, the long distance between fibers causes small angular drifts in the beam to manifest as large translations of the beam incident on the second fiber, and therefore large coupling efficiency falloff. The relay imaging lenses suppress this effect.}
%ith the relay imaging lenses, the dependence of the lateral position of the beam by the second fiber on the angular drifts by the first fiber is suppressed.}%lenses reduce effective lever arm?
\end{centering}
\end{figure}

The delivery optics will drift on hour timescales in response to temperature changes in the area, and fluctuate on faster timescales in response to seismic and anthropogenic vibrations in the environment, along with air currents. To suppress the influence of pointing fluctuations on the fiber coupling efficiency, we opted to add two long focal length lenses in a `4f' relay-imaging configuration to better passively isolate the coupling efficiency of the interferometer beam against pointing fluctuations of the steering optics, effectively reducing the lever arm associated with the beam propagation (see Fig. \ref{fig:relay_imaging_cartoon}).  Moreover, the interferometer beam propagates in rough vacuum ($\approx 1 \text{ Torr}$) to suppress the influence of air currents on the pointing of the beam. We tested the efficacy of using a relay imaging system to stabilize the fiber coupling efficiency in a full-scale prototype of the laser transport system (see Sec. \ref{sec:experimental_test_horizontal_config} for further details).

\begin{figure}%[H]%[hbt!]
\begin{centering}
\includegraphics[width=3.37in]{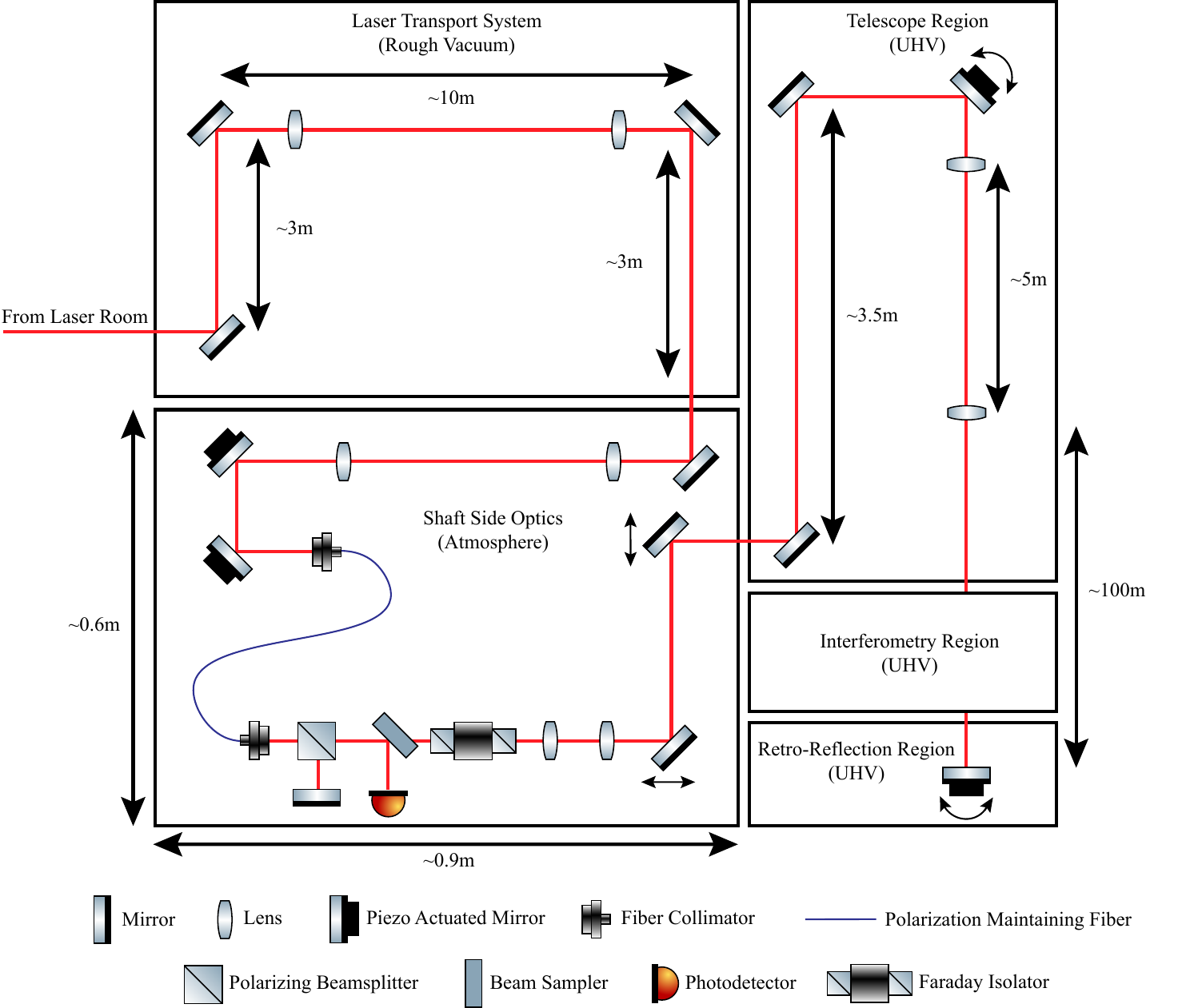}
\caption{\label{fig:beam_delivery_system_cartoon} Optical schematic of the full beam delivery system.  See main text for details.}
\end{centering}
\end{figure}

\begin{figure}%[H]%[hbt!]
\begin{centering}
\includegraphics[width=3.37in]{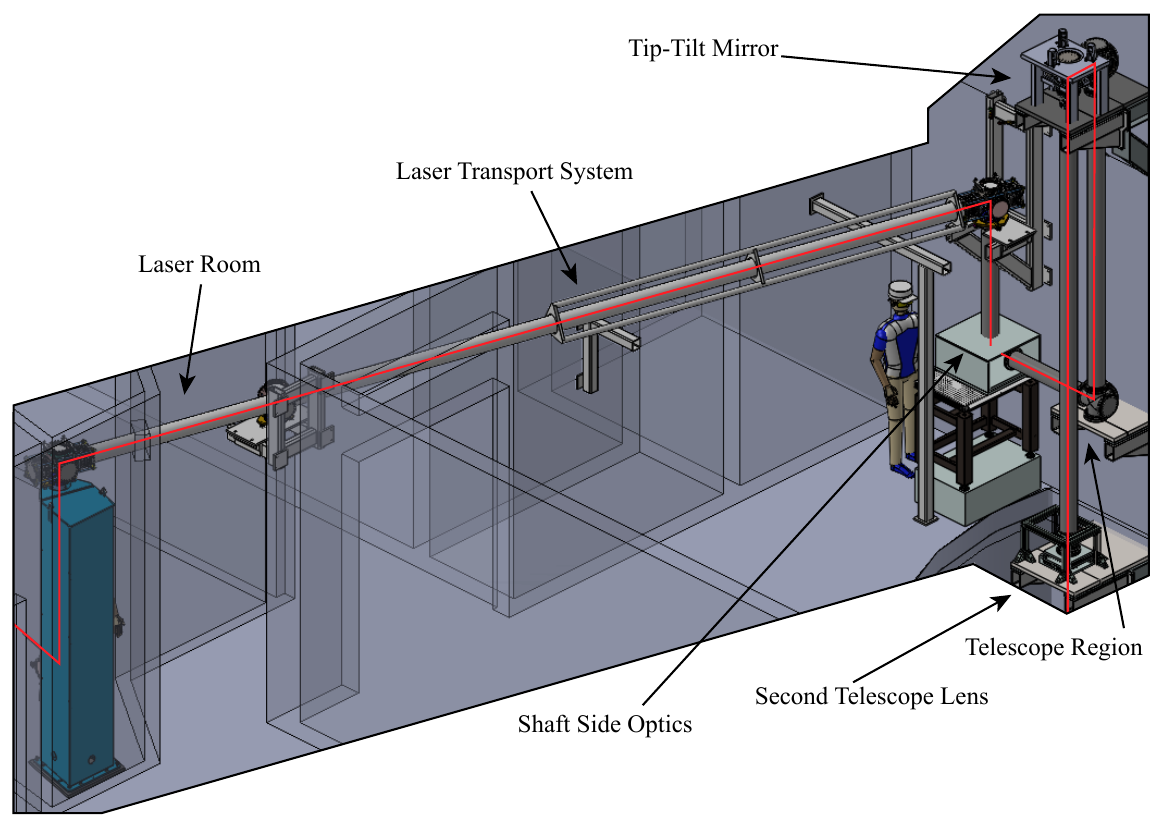}
\caption{\label{fig:beam_delivery_system_CAD} CAD model of the beam delivery system whose optical configuration is illustrated in Figure \ref{fig:beam_delivery_system_cartoon}.}
\end{centering}
\end{figure}
%this figure was edited in 2023-9-26 (beam delivery system mechanical design graphic) on the northwestern desktop

%CAD model from Jeff is in MAGIS Dropbox\Kovachy Group\CAD\MAGIS\Beam delivery\v1.0\from Jeff\LTS Brackets 8_1_23
%lots of missing cross chambers
%lots of missing other components
%unorganized organization structure
%very large file
%might take a full day to generate a really nice figure -- what's the minimum we can do here?

%things maybe we'd need to do on the CAD front
%no laser tables -- nothing really at all in the laser room
%missing cross chambers in both balanced bellows configurations
%no thermionics stage
%old telescope region model which has vertical shaft
%superimposed new telescope region without second telescope area
%optics box missing cover on one side

%concrete slab -- w/ or w/out
%where to indicate the telescope lenses are?
%have a retro prism as part of the LTS or not?
%old model of bottom lens translation stage

%paragarph on the shaft-side optics
The `shaft-side optics' (see Figures \ref{fig:beam_delivery_system_cartoon} and \ref{fig:beam_delivery_system_CAD}) consist of the short fiber described above, which provides initial spatial filtering of the beam \cite{MAGIS-100:2021etm}, a photodetector for power monitoring (used for power stabilization feedback as described above), optics for a fiber-phase-noise-cancellation setup, and a short telescope. One fixed mirror on the shaft-side breadboard will direct the interferometer beam to a short telescope, which shapes the beam to match the mode of the short fiber.
%through a viewport window which separates rough vacuum from the shaft-side optics in atmosphere.  This mirror directs the beam through a short telescope, shapes the beam to match the mode of the short fiber to improve the fiber coupling efficiency.
Two subsequent piezo-actuated mirror mounts adjust the angle and lateral position of the interferometer beam into the fiber. These mounts can be adjusted every $\approx 40 \text{ hrs}$ to compensate for long timescale drifts in the laser transport system alignment. The polarizing beamsplitter and mirror just after the fiber back-reflect a small portion of the interferometer beam as part of a fiber-phase-noise-cancellation scheme \cite{DeRose:23,Ma:94, MAGIS-100:2021etm}. A subsequent Faraday isolator keeps preceding optics protected from back-reflected light, and then a short telescope shapes the interferometer beam so that it is focused to a radial waist of $\approx 300\;\mu\text{m}$ at the location of the tip-tilt mirror, $\approx 4.5\text{ m}$ away. Two subsequent mirrors mounted to motorized translation stages provide fine adjustment of the lateral alignment of the interferometer beam into the telescope region. Adjusting the lateral position of the interferometer beam may also be useful to keep the beam centered on atoms for extended LMT sequences when the effective mirror and beamsplitter pulses take a non-negligible amount of time.

%paragarph on the telescope region
The `telescope region' (see Figure \ref{fig:beam_delivery_system_cartoon}) exists in UHV ($10^{-9}\text{ Torr}$) and consists of $\approx 4.5\text{m}$ of initial free propagation (1\;m of  horizontal propagation and 3.5\;m of vertical), which further cleans the spatial mode of the beam\cite{MAGIS-100:2021etm}. The beam then reflects off the tip-tilt mirror, whose mechanical configuration is discussed earlier in Section \ref{sec:tip_tilt_system_mechanical_design}, and travels through the $M=30$ telescope discussed in section \ref{sec:spherical_lens_aberrations} which expands the interferometer beam to a waist of $\approx 1 \text{ cm}$. The piezo-actuated mirror stage at the bottom of the shaft \cite{MAGIS-100:2021etm} retro-reflects the interferometer beam back up the shaft. The telescope de-magnifies angles by a factor equal to its magnification $M$, reducing the impact of angular fluctuations of pre-telescope optics.

\begin{comment}
\begin{figure}[H]%[hbt!]
\begin{centering}
\includegraphics[width=3.0in]{laser_transport_system_figures/introduction_figures/sub_component_diagram.png}
\caption{\label{fig:subcomponent_cartoon} subcomponents of beam delivery system}
\end{centering}
\end{figure}
%inset on a figure
%adobe illustrator -- export as pdf
%   pdfs are vector graphics (?) -- some types
% give us nice resolution
\end{comment}

\subsection{\label{sec:experimental_test_horizontal_config}Full Scale Prototype of Laser Transport System}%Experimental Test of 4f Relay Imaging System Over 20m Baseline

\begin{figure}[H]%[hbt!]
\begin{centering}
\includegraphics[width=3.37in]{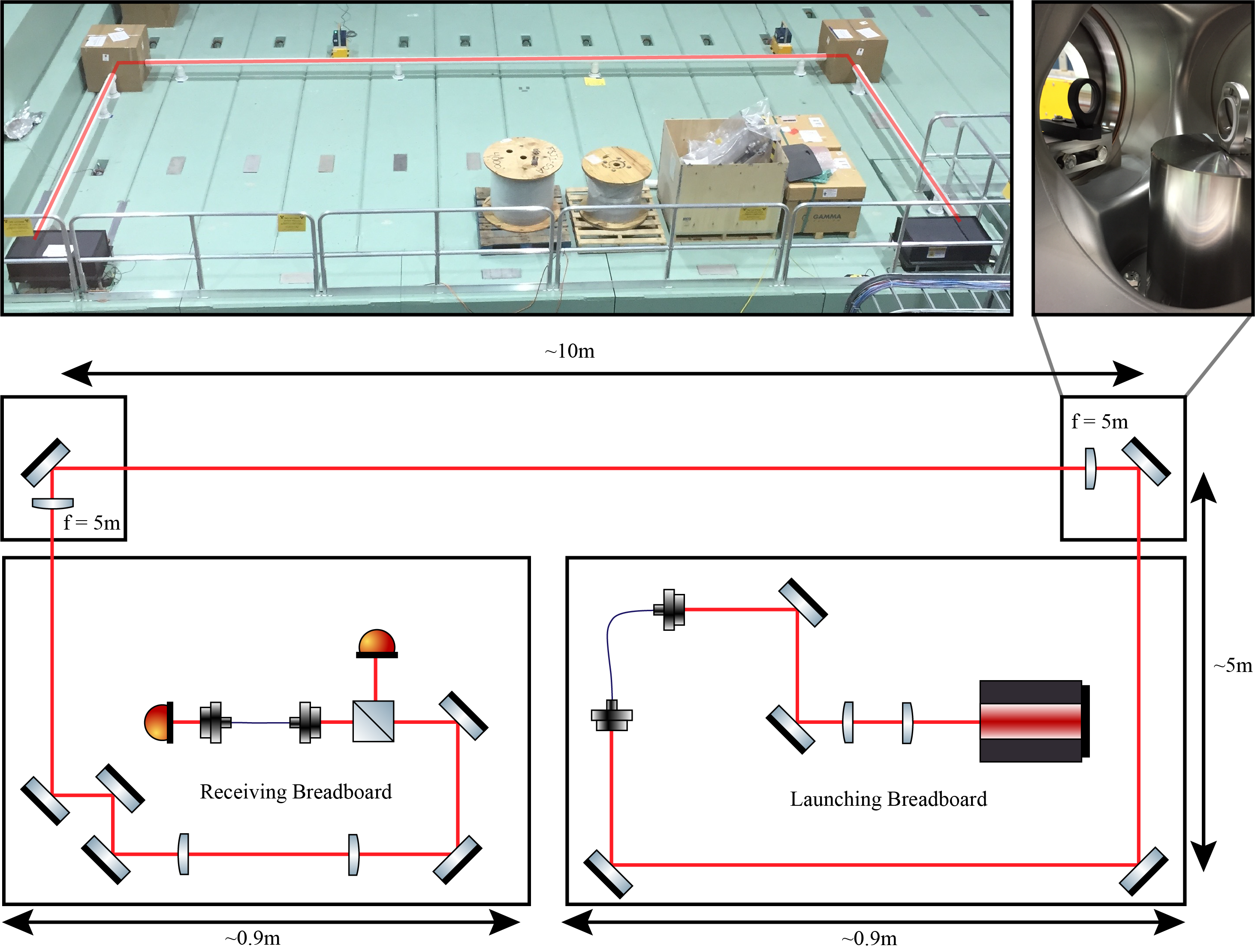}
\caption{\label{fig:horziontal_4f_at_nml_picture}An image and optical schematic of the prototype test of the laser transport system.  See main text for details.}
\end{centering}
\end{figure}

\begin{figure}[H]%[hbt!]
\begin{centering}
\includegraphics[width=3.37in]{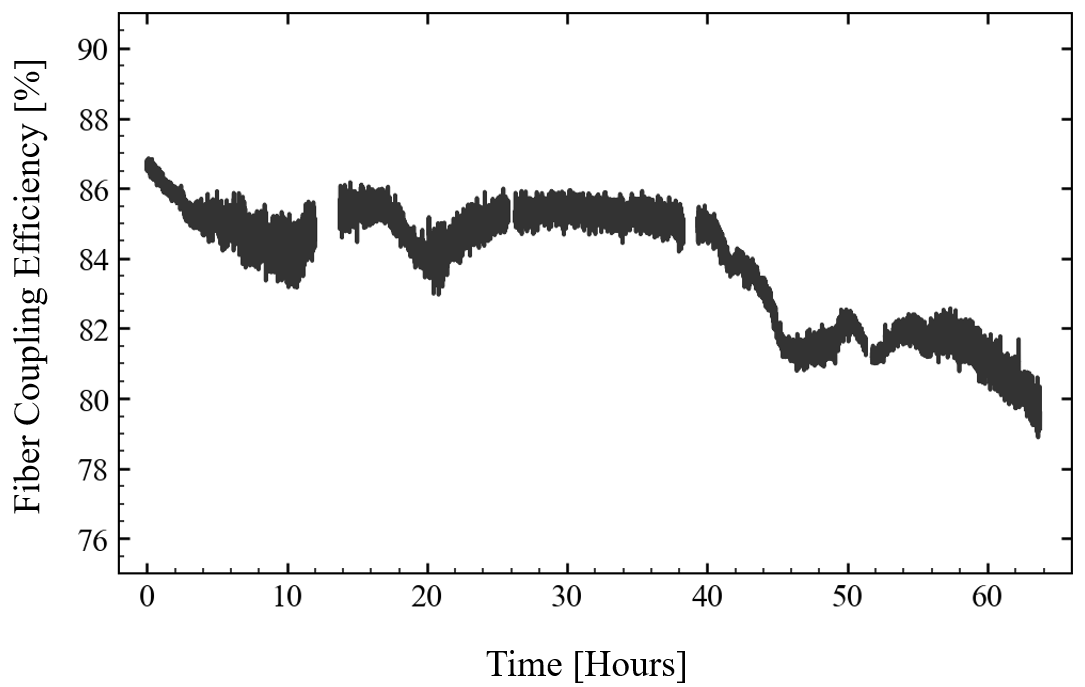}
\caption{\label{fig:nml_time_trace} Trace of the fiber coupling efficiency as a function of time. The raw data was sampled at a rate of 1\;kHz, and the data as presented here is averaged over 10 second bins.}
\end{centering}
\end{figure}

%introduction to the experiment
We constructed a prototype of the laser transport system to demonstrate the feasibility of coupling the interferometer beam into a short fiber following $\approx 20 \text{ m}$ of free propagation
%its propagation through the laser transport system
%, consequently reducing the susceptibility of the interferometer beam's pointing stability to environmental perturbations,
and to monitor the passive stability of the fiber coupling efficiency over time scales of many hours in the presence of environmental perturbations. This test was performed in a horizontal configuration (see Fig. \ref{fig:horziontal_4f_at_nml_picture}) on the rooftop of the Fermilab Accelerator Science and Technology accelerator in the New Muon Lab on the Fermilab campus and included the right-angle turns that will be necessary to elevate the beam off the floor in the final configuration.
%, resulting in a total optical path length of $\approx 20 \text{ m}$.
The relay imaging system was composed of two f = 5\;m lenses in a `4f' configuration respectively mounted to two vacuum chambers which each housed a fixed right angle mirror. This proof of concept was without any of the vertical support structures planned for the full laser transport system, since those structures were still being designed. As shown in Figure \ref{fig:nml_time_trace}, over a continuous 40-hour period, we successfully maintained a $>80 \%$ fiber coupling efficiency without the need for any alignment adjustments.

%this figure was composed in 2023-9-25 (NML horizontal test setup graphic)

The prototype consisted of a `launching' breadboard which consisted of a $\approx 5 \text{ mW}$ laser (CivilLaser) which was coupled into a fiber to clean the spatial mode of the beam, and directed to a right angle mirror and 5\;m focal length lens (EKSMA 110-0289E) $\approx 5\;\text{m}$ away. The beam then propagated $\approx 10$\;m, before reflecting off a second right angle mirror paired with a second 5\;m focal length lens. Subsequently, the beam propagated over a $\approx 5 \text{ m}$ distance to the `receiving' breadboard. On this board, three mirrors set the alignment of the beam into a short telescope which mode-matched the beam into a short fiber. A 50:50 beamsplitter and photodiode combination just before the fiber monitored the power of the beam before the fiber, and a photodiode after the fiber measured the post-fiber power. To extract a coupling efficiency, we took the ratio of the real-time power before and after the fiber so that the measured coupling efficiency was independent of power fluctuations in the laser. 

We obtained best stability results when using Thorlabs Polaris mirror mounts (POLARIS-K1S4) with the pitch and yaw lead screw axes locked with locking nuts. To suppress the higher frequency jitter induced by air currents, the beam was enclosed by PVC pipe in the long regions of free-space propagation, and cardboard boxes with two holes cut in them served as local enclosures for the right angle mirrors. Both the launching and receiving breadboards were enclosed with hardboard enclosure material from Thorlabs. While air current shielding did not noticeably affect the long term stability, it slightly suppressed root mean square (rms) fluctuations from $0.5 \%$ to $0.4\%$ integrated over a 20 mHz to 4 kHz band. As shown in Figure \ref{fig:nml_time_trace}, we measured a $>80\text{\%}$ fiber coupling efficiency over a $40 \text{ hr}$ period despite this $\approx 20 \text{ m}$ path length. To compensate for drifts on these large timescales, automated tweaks of fiber coupling with piezo-controlled mirror mounts, indicated in Figure \ref{fig:beam_delivery_system_cartoon}, will be used.
%air currrent shielding improvement numbers came from D:/2023-3-3 (traces from fermilab)/NML_second_test and NML_third_test and data was analysed in plotting_traces.ipynb

In the near future, we plan to test the stability in the final vertical configuration to evaluate the impacts of the vertical support structures.

\begin{comment}
This test was done in atmosphere, and although we did have some air current shielding, we expect the high frequency fluctuations from air currents to be further suppressed in the full system. It's possible also in the vertical configuration the vibration effects are going to be a bit larger since the right angle mirrors will be mounted to taller towers with larger mechanical response to vibrations. OTHER THINGS WE COULD SAY? Mention that the concrete blocks were moving? We don't have data arguing for that having an influence?
\end{comment}

% to tell a simpler story, where we're just changin one variable

% unlocked polaris mounts
% flexture mounts
% locked polaris mounts

%adding air currently shielding reduced the short term rms from X% to Y%

% something we notices that made a difference was the type of mount we used

% if we don't definitely know something -- we can just show the black trace
    % just say, we were able to get the performance to this level

% go back and see air current shielding

%notes from Kovachy
%show the end result — over a couple days it stays pretty stable
%whatever we can say definitively about causes of stability, we can try to say that
%Time traces — long timescales, with and without the locked Polaris (maybe) bare minimum, where we show the trace and it was good. For the air currently shielding, if we do go back and find that it made a difference. we could look at a zoomed in time trace and see if the air currents made a difference. could do an fft and see if the air currents make much of a difference

\section{Conclusion}
In this paper, we introduced a method for Coriolis force compensation in long-baseline atom interferometers, provided specific implementation details for MAGIS-100, and presented the design of the beam delivery system for MAGIS-100. In the future, we will demonstrate the pivot point method experimentally in MAGIS-100 and further evaluate the feasibility of using this method for future km-baseline interferometers.

\begin{comment}
In this paper, we have introduced and described the pivot point method for Coriolis force compensation for long baseline atom interferometers, which consists of pivoting the interferometer beam about a point on the interferometer axis and tuning the initial kinematics of the atom cloud so that the atoms are centered in the laser beam for the beam splitter and mirror light-atom interactions. We have presented example experiment cycles for performing dual-isotope atom interfeormetery with a single atom source over a full $\approx 100 \text{ m}$ baseline, as well as a for the case of performing atom gradiometery over a $\approx 100 \text{ m}$ baseline. We have discussed how the position of the pivot point can be adjusted, and presented the mechanical design of the tip tilt system that will be used for MAGIS-100. We have numerically computed the scale of the phase aberration induced on the interferometer beam by the spherical nature of the telescope lenses, and discuss challenges associated with using this method for even longer interferometer baselines. In addition, we have presented the mechanical design of the beam delivery system for MAGIS-100 and present data from the prototype test of the laser transport system.
\end{comment}

%\section{Acknowledgments}
\begin{acknowledgments}

We thank Jason Hogan and James Santucci for valuable discussions and technical contributions. This project is funded in part by the Gordon and Betty Moore Foundation Grant GBMF7945. Some of the work presented in this document leveraged the resources of the Fermi National Accelerator Laboratory (Fermilab), a U.S. Department of Energy, Office of Science, HEP User Facility. Fermilab is managed by Fermi Research Alliance, LLC (FRA), acting under Contract No. DE-AC02-07CH11359.  This work is supported in part by the U.S. Department of Energy, Office of Science, QuantiSED Intitiative. We also acknowledge support from the David and Lucile Packard Foundation (Fellowship for Science and Engineering).
\end{acknowledgments}

\section*{Author Declarations}

\noindent \textbf{Conflict of Interest}

\noindent The authors have no conflicts to disclose.

\vspace{5mm}

\noindent \textbf{Author Contributions}

\vspace{5mm}

\noindent \textbf{Jonah Glick:} Conceptualization (equal), Methodology (equal), Investigation (lead), Validation (equal), Formal Analysis (lead),  Writing -- original draft (lead). Writing -- review \& editing (equal). \textbf{Zilin Chen:} Methodology (equal), Investigation (supporting), Validation (equal), Writing -- review \& editing (supporting). \textbf{Tejas Deshpande:} Conceptualization (equal), Methodology (equal), Investigation (supporting), Validation (equal), Supervision (equal), Writing -- review \& editing (supporting). \textbf{Yiping Wang:} Conceptualization (equal), Methodology (equal), Investigation (supporting), Validation (equal), Writing -- review \& editing (supporting). \textbf{Tim Kovachy:} Conceptualization (equal), Methodology (equal), Validation (equal), Supervision (equal), Writing -- review \& editing (equal).

\section*{Data Availability Statement}

\noindent The data that support the findings of this study are available from the corresponding author upon reasonable request.

\section*{Conflict of Interest Statement}
\noindent The authors have no conflicts to disclose.

\bibliography{bibliography}% Produces the bibliography via BibTeX.

\end{document}